\documentclass[twocolumn,10pt]{autart}    

\usepackage{graphicx}          
\usepackage{epstopdf}
\usepackage{setspace}                          
\usepackage{amsmath}
\usepackage{enumerate}
\usepackage{graphicx}          
\usepackage{tikz}
\usepackage{cite}
\usepackage{natbib}                          

\usepackage{amsfonts}
\usepackage{algorithm}
\usepackage{algpseudocode}
\usepackage{amssymb}
\usepackage{subfigure}
\usepackage{hyperref}
\hypersetup{
    colorlinks = true,
    citecolor = cyan,
}

\usepackage{wrapfig}
\usepackage{color}
\def\({\left(}
\def\){\right)}
\DeclareMathOperator{\sign}{sign}
\DeclareMathOperator{\diag}{diag}
\DeclareMathOperator{\col}{col}

\newtheorem{theorem}{Theorem}[section]

\newtheorem{lemma}{Lemma}[section]

\newtheorem{assumption}{\bf Assumption}
\newtheorem{rema}{Remark}[section]

\begin{document}
\begin{frontmatter}
\title{Distributed Estimation with Quantized Measurements and Communication over Markovian Switching Topologies\thanksref{footnoteinfo}}
\thanks[footnoteinfo]{The work is supported by National Natural Science Foundation of China under Grants 62025306, 62433020, 62303452, T2293770 and 62473040, and CAS Project for Young Scientists in Basic Research under Grant YSBR-008.
The material in this paper was not presented at any conference.
Corresponding author:  Yanlong Zhao.}

\author[AMSS,KTH]{Ying Wang}\ead{wangying96@amss.ac.cn},
\author[AMSS]{Jian Guo}\ead{j.guo@amss.ac.cn},
\author[AMSS,UCAS]{Yanlong Zhao}\ead{ylzhao@amss.ac.cn},
\author[ZYUT,AMSS,UCAS]{Ji-Feng Zhang}\ead{jif@iss.ac.cn}
\address[AMSS]{Key Laboratory of Systems and Control, Academy of Mathematics and Systems Science, Chinese Academy of Sciences, Beijing 100190, P. R. China.}
\address[KTH]{Division of Decision and Control Systems, KTH Royal Institute of Technology, Stockholm 11428, Sweden.}
\address[UCAS]{School of Mathematical Sciences, University of Chinese Academy of Sciences, Beijing 100049, P. R. China.}
\address[ZYUT]{School of Automation and Electrical Engineering, Zhongyuan University of Technology, Zhengzhou 450007, P. R. China}

\begin{abstract}
This paper addresses distributed parameter estimation in stochastic dynamic systems with quantized measurements, constrained by quantized communication and Markovian switching directed topologies. To enable accurate recovery of the original signal from quantized communication signal, a persistent excitation-compliant linear compression encoding method is introduced.
Leveraging this encoding, this paper proposes an estimation-fusion type quantized distributed identification algorithm under a stochastic approximation framework. The algorithm operates in two phases: first, it estimates neighboring estimates using quantized communication information, then it creates a fusion estimate by combining these estimates through a consensus-based distributed stochastic approximation approach.
To tackle the difficulty caused by the coupling between these two estimates, two combined Lyapunov functions are constructed to analyze the convergence performance.
Specifically, the mean-square convergence of the estimates is established under a conditional expectation-type cooperative excitation condition and the union topology containing a spanning tree. Besides, the convergence rate is derived to match the step size's order under suitable step-size coefficients. Furthermore, the impact of communication uncertainties including stochastic communication noise and Markov-switching rate is analyzed on the convergence rate.
A numerical example illustrates the theoretical findings and highlights the joint effect of sensors under quantized communication.
\end{abstract}
\end{frontmatter}

\section{Introduction}
\vspace{-5pt}
\subsection{Background and Motivations}
\vspace{-5pt}
Wireless sensor networks (WSNs) are extensively used in fields such as environmental monitoring, intelligent transportation, and smart agriculture due to their easy deployment, flexibility, low power consumption, high accuracy, and reliability \citep{ASSC2002,TC2011,RSCB2014,GHZDY2020}. Distributed parameter estimation over sensor networks is a critical theoretical issue in WSNs research and has garnered significant academic attention \citep{DKMRS2010,BFH2013,MG2012}. Various effective distributed estimation algorithms have been developed, including distributed stochastic approximation algorithms \citep{ZZ2012,LC2020}, distributed stochastic gradient algorithms \citep{GL2022,SSP2020}, and distributed least squares algorithms \citep{TYS2010,XZG2021}. These distributed approaches only rely on partial information and local data exchange due to collaboration constraints caused by the sensing and communication limitations of individual sensors. However, it is important to note that these algorithms generally assume both accurate communication and precise measurements.

Actually, WSNs consists of numerous spatially distributed sensor nodes, each powered by batteries and equipped with sensing, communication, and computational capabilities. Consequently, WSNs face significant energy and communication constraints due to limited battery life in addition to the inherent collaboration limitations \citep{ASSC2002,RSCB2014}. Key constraints include:
\vspace{-5pt}
\begin{itemize}
\item Energy constraints \citep{HWDYA2023,HWDYA2023I}: Limited sensing and measurement ranges, restricted processing capabilities, small memory, and more.
\item Communication constraints \citep{GHZDY2020}: Short communication bandwidth, time-varying network topology, network link failures, and more.
\end{itemize}
\vspace{-5pt}
It is well-known that both data transmission and measurement consume more energy than local data processing, with communication being the most energy-intensive of the three \citep{PK2020}. Therefore, reducing the bandwidth required for data transmission and the precision of measurement data can significantly extend battery life. Consequently, designing distributed adaptive estimation algorithms that can cooperatively estimate unknown parameters based on quantized transmitted data and quantized measurement data has become an increasingly important research focus.

\vspace{-5pt}
\subsection{Related literature}
\vspace{-5pt}
Quantized data refers to data where only the set it belongs to is known, without precise knowledge of exact values, i.e., data with finite precision or bits \citep{WZY2003}. The existing works on distributed estimation with quantized data generally falls into two categories:

\textbf{Distributed estimation with quantized measurements}: Here, each sensor has quantized measurements but can communicate accurately with its neighbors \citep{DWDG2017,WZ2019,WZZ2021,FCZ2022,HWDYA2023,HWDYA2023I}. For example, \cite{FCZ2022} investigated distributed estimation with binary-valued measurements in time-varying networks, proposing a distributed stochastic approximation algorithm with extended truncation and proving its almost sure convergence under strictly stationary signals. Besides, \cite{WZZ2021} developed a Quasi-Newton type projection algorithm with a time-varying projection operator and a diffusion strategy under binary measurements and an undirected, connected topology, achieving almost sure convergence without relying on signal periodicity, independence, or stationarity.

\textbf{Distributed estimation with quantized communication}: In this approach, sensors have accurate measurements but can only communicate with neighbors through quantized channels \citep{KMR2012,ZSX2015,ZLSX2017,ZCGXLJ2018,LWLX2023}. For instance, \cite{KMR2012} proposed a consensus+innovations distributed algorithm based on stochastic approximation idea and probabilistic quantizer-based communication, showing almost sure convergence in undirected and connected topologies. Additionally, \cite{ZSX2015} and \cite{ZCGXLJ2018} introduced running average distributed estimation algorithms within the consensus+innovation framework, using probabilistic quantized communication with only one measurement per sensor.

While these approaches are insightful, the first class addresses quantized measurement alone, without considering the complexity and energy demands of quantized communication. The second class, meanwhile, overlooks the irreversible loss of information caused by quantized measurements. Furthermore, both approaches fail to account for the effects of unreliable communication networks, such as time-varying topologies and stochastic communication noise.
The goal of this paper, therefore, is to develop a distributed estimation algorithm that integrates both quantized measurements and quantized communication under Markov-switching topologies and stochastic communication noise, thereby reducing communication bandwidth and measurement precision requirements to extend sensor battery life substantially.

\vspace{-5pt}
\subsection{Main contribution}
\vspace{-5pt}
This paper investigates the distributed parameter estimation problem with both quantized measurements and quantized communication in the presence of Markov-switching directed topologies and stochastic communication noise. The primary contributions of this study, in contrast to previous work, are as follows:
\begin{itemize}

\item \textbf{Problem aspect}: This work is the first to address distributed parameter estimation with both quantized measurements and quantized communication, while also accounting for communication uncertainties such as Markov-switching topologies and stochastic communication noise \citep{KMR2012,ZSX2015,ZCGXLJ2018,LWLX2023,DWDG2017,WZZ2021,FCZ2022}. Additionally, it achieves one-bit communication through a linear compression encoding method, significantly reducing communication costs compared to prior methods \citep{KMR2012,ZSX2015,ZCGXLJ2018,LWLX2023}.

\item \textbf{Algorithm aspect}: An estimation-fusion type quantized distributed identification (EFTQDI) algorithm is proposed to handle quantized measurements and quantized communication. To counteract the information reduction and nonlinearity from quantized communication, the estimate of neighboring estimate is designed using a stochastic approximation approach. Notably, a linear communication encoding method satisfying persistent excitation condition \citep{WZZG2022} is developed, ensuring that the original information can be reconstructed from one-bit transmitted data. Then, the algorithm creates a fusion estimate by using stochastic approximation idea through a combination of these estimates and a consensus strategy.

\item \textbf{Result aspect}:  By constructing two Lyapunov functions to analyse the coupling relationship of two estimates, this paper establishes the mean-square convergence of both the fusion estimate and the estimate of neighboring estimate under a conditional expectation-type cooperative excitation condition, which is more general than typical cooperative excitation conditions \citep{KMR2012,ZZ2012,FCZ2022}. Additionally, the mean-square convergence rate of the EFTQDI algorithm is shown to match the order of the step size. This work further analyzes the impact of communication uncertainties, such as stochastic communication noise and Markov-switching rates, on the convergence rate.
\end{itemize}

\vspace{-5pt}
The remainder of this paper is organized as follows. Section \ref{sec:PF} introduces graph preliminaries and problem formulation. Section \ref{sec:EFQDIA} presents the proposed algorithm, while Section \ref{sec:conver} demonstrates its convergence properties. Proofs of the main results are provided in Section \ref{sec:proof}. Section \ref{sec:sim} includes a numerical example to illustrate the main findings, and Section \ref{sec:con} concludes the paper and discusses potential future work.

\vspace{-5pt}
\noindent\emph{Notations.} This paper uses $x=[x_1,\ldots,x_n]^T\in\mathbb{R}^n$ and $A\in\mathbb{R}^{m\times n}$ to denote $n$-dimensional vector and $m\times n$-dimensional real matrix respectively. Moreover, we denote $\|x\|=\sum_{i=1}^n{x_i^2}$ and $\|A\|=\left(\lambda_{\max}(A^TA)\right)^{\frac{1}{2}}$ as the Euclidean norm of vector and matrix respectively, where the notation $T$ denotes the transpose operator and $\lambda_{\max}(\cdot)$ denotes the largest eigenvalue of the matrix. Correspondingly, we use $\lambda_{\min}(\cdot)$ to denote the smallest eigenvalue of the matrix.
The function $I_{\{\cdot\}}$ denotes the indicator function, whose value is 1 if its argument  is true, and 0, otherwise.
Let $A$ and $B$ be two symmetric matrices, then $A\geq B$ means that $A-B$ is a positive semi-definite matrix. The Kronecker product of two matrices $A=\{a_{ij}\}\in\mathbb{R}^{m\times n}$ and $B\in\mathbb{R}^{p\times q}$ is defined as
$$A\otimes B=\left(
               \begin{array}{ccc}
                 a_{11}B & \cdots & a_{1n}B \\
                 \vdots & \ddots & \vdots \\
                 a_{m1}B & \cdots & a_{mn}B \\
               \end{array}
             \right)\in\mathbb{R}^{mp\times nq}.
$$

\vspace{-5pt}

\section{Problem formulation}\label{sec:PF}
\subsection{Graph Preliminaries}
In order to describe the relationship between sensor nodes, a sequence of time-varying digraphs $\mathcal{G}_k= (\mathcal{V}, \mathcal{E}_k,\mathcal{A}_k)$ are introduced here, where $\mathcal{V}=\{1,2,\ldots,m\}$ is the set of nodes, and the node $i\in\mathcal{V}$ represent the sensor node $i$.
$\mathcal{E}_k\subseteq\mathcal{V}\times\mathcal{V}$ is the edge set which is used to describe the communication between agents at time $k$, and a directed edge $(i,j)\in \mathcal{E}_k$ if and only if there is a communication link from $i$ to $j$ at time $k$.
$\mathcal{A}_k=\{a_{ij,k}\}\in\mathbb{R}^{m\times m}$ is the adjacency matrix of $\mathcal{G}_k$. The elements of the matrix $\mathcal{A}_k$ satisfy: $0<a_{ij,k}<1$ if and only if $(j,i)\in \mathcal{E}_k$, otherwise $a_{ij,k}=0$.
The set of the neighbors of agent $i$ at time $k$ is denoted as $\mathcal{N}_{i,k}=\{j\in \mathcal{V}|(j,i)\in\mathcal{E}_k\}$ and each agent can only exchange information with its neighbors.
The Laplacian matrix of $\mathcal{G}_k$ is defined as $\mathcal{L}_k=\diag\{\sum_{j=1}^{m}a_{1j,k},\sum_{j=1}^{m}a_{2j,k},\ldots,\sum_{j=1}^{m}a_{mj,k}\}-\mathcal{A}_k$.

\vspace{-5pt}
The dynamic changes of the communication topology digraph $\mathcal{G}_k$ is described with a Markovian chain $\{m_k,k=1,2,\ldots\}$ with a finite state space
$\mathcal{S}=\{1,2,\ldots,s\}$ and transition probability $p_{uv}=\mathbb{P}\left(m_{k+1}=v \mid m_k=u\right)$. Then the corresponding communication topology digraph set is $\{\mathcal{G}^{(1)}, \ldots, \mathcal{G}^{(s)}\}$, where  $\mathcal{G}^{(u)}=(\mathcal{V}, \mathcal{E}^{(u)}, \mathcal{A}^{(u)})$ is a digraph, $\mathcal{E}^{(u)}$, $\mathcal{A}^{(u)}=\{a_{ij}^{(u)}\}$ and $\mathcal{L}^{(u)}$ are the corresponding edge set, adjacency matrix and Laplacian matrix for $u\in S$. We say $m_k=u$ if and only if $\mathcal{G}_k =\mathcal{G}^{(u)}$.
The union topology of $\mathcal{G}^{(1)}, \ldots, \mathcal{G}^{(s)}$ is denoted by $\mathcal{G}_{\mathcal{C}}=\(\mathcal{V},\mathcal{E}_{\mathcal{C}}, \mathcal{A}_{\mathcal{C}}\)$, where $\mathcal{E}_{\mathcal{C}}=\bigcup_{u=1}^s\mathcal{E}^{(u)}$ and $\mathcal{A}_{\mathcal{C}}=\left\{{a}_{ij}^{\mathcal{C}}\right\}$.
A directed tree is a digraph, where each node except the root has exactly one parent node.
A spanning tree of $\mathcal{G}_{\mathcal{C}}$ is a directed tree whose node set is $\mathcal{V}$ and whose edge set is a subset of $\mathcal{E}_{\mathcal{C}}$.
Besides, $\mathcal{G}_{\mathcal{C}}$ is called a balanced digraph if $\sum_{j=1}^m a_{ij}^{\mathcal{C}}=\sum_{j=1}^m a_{ji}^{\mathcal{C}}$ for all $i\in \mathcal{V}$.
$\mathcal{G}_{\mathcal{C}}$ is called an undirected graph, if $\mathcal{A}_{\mathcal{C}}$ is symmetric.
The mirror graph $\hat{\mathcal{G}}_{\mathcal{C}}$ of the digraph $\mathcal{G}_{\mathcal{C}}$ is an undirected graph, denoted by $\hat{\mathcal{G}}_{\mathcal{C}}=(\mathcal{V},\hat{\mathcal{E}}_{\mathcal{C}}, \hat{\mathcal{A}}_{\mathcal{C}})$ with $\hat{\mathcal{A}}_{\mathcal{C}}=\{\hat{a}_{ij}^{\mathcal{C}}\}$, $\hat{a}_{ij}^{\mathcal{C}} =\hat{a}_{ji}^{\mathcal{C}}=({a}_{ij}^{\mathcal{C}}+{a}_{ji}^{\mathcal{C}})/2$.

\vspace{-5pt}
\begin{rema}\label{RGP}
The switching topology described by the above Markovian chain can effectively model link failures or losses between the channels of a WSN \citep{ZZ2012}.
Actually, the number of failure-prone links $m_f$ is finite with $m_f\leq m(m-1)$. It implies that the total possible configurations of the WSN with link failures is also finite, with an upper bound of $2^{m_f}$.
We use the Markovian chain $m_k$ to represent these link failures at time $k$, where $m_k$ is an $m_f$-dimensional vector with each component taking a value of $0$ or $1$ to indicate whether a link is absent or present.
In this case, the state space of the Markovian chain $m_k$ is finite and denoted by $\mathcal{S}$. The corresponding topology are represented by $\{\mathcal{G}^{(1)}, \ldots, \mathcal{G}^{(s)}\}$, meaning that the Markov-switching topology can describes various link failure scenarios of WSNs.
In addition, the Markovian chain model can also capture the uncertain aspects of packet loss in digital communications \citep{HDNM2010}.
\end{rema}

\subsection{System Description}
Consider a multi-agent network consisting of $m$ sensors, the dynamic of the $i$th $(i=1,\ldots,m)$ sensor at $k$-th times is a linear stochastic model with quantized measurements as follows:
\begin{gather}\label{M}
\left\{
   \begin{array}{ll}
     y_{k,i}=\phi_{k,i}^T\theta+d_{k,i},\\
     s_{k,i}=I_{\{y_{k,i}\leq C_i\}},
   \end{array}
 \right.
\end{gather}
where $\phi_{k,i}\in \mathbb{R}^n$ is an $n$-dimensional regressor, $d_{k,i}$ is a stochastic noise and $\theta\in\mathbb{R}^n$ is an unknown parameter vector to be estimated. And $y_{k,i}$ is a scalar measurement of sensor $i$, which cannot be exactly measured. What can be measured is only the quantized information $s_{k,i}$, where $C_i$ is the fixed threshold of the binary sensor $i$.

\vspace{-5pt}

The information communication between two adjacent sensors is as follows: at the sender side of the communication channel $(j,i)\in\mathcal{E}_k$, the sensor $j$ sends its estimate $\theta_{k,j}$ to the sensor $i$. Due to the uncertainties of communication channels, at the receiver side of the channel $(j,i)$, the sensor $i$ receives the encoding value of an estimate of $\theta_{k-1,j}$ as
\vspace{-10pt}
\begin{equation}\label{est_c}
\left\{
\begin{aligned}
&y_{k,ij}=Q(\theta_{k-1,j})+\omega_{k,ij},\\
&z_{k,ij}=I_{\{y_{k,ij}\leq C_{ij}\}},
\end{aligned}
\right.
\end{equation}
where $Q(\cdot):\mathbb{R}^n\rightarrow\mathbb{R}$ is a compression encoding function to be designed, which could transform the vector to the scalar;
$\omega_{k,ij}$ is the stochastic additive communication noise, which can be used to model the thermal noise and channel fading \citep{LZ2010}; $z_{k,ij}$ is the one-bit quantized data that the sensor $i$ collects from its neighbor $j$, where $C_{ij}$ is the fixed threshold of the channel $(j,i)$.

\vspace{-10pt}
\subsection{Assumptions}
\vspace{-5pt}
In order to proceed our analysis, we introduce some assumptions concerning the topologies, regressors, priori information of the unknown parameter and noises. Define the $\sigma$-algebra $\mathcal{F}_k=\{\theta_{0,i}, m_{l}, d_{l,i},\phi_{l+1,i}, \omega_{l,ij} \text{ for } i\in \mathcal{V} \text{ and } j\in \mathcal{N}_{i,l}, 0\leq l\leq k\}$.
\vspace{-5pt}
\begin{assumption}\label{AG}
The digraph $\mathcal{G}^{(u)}$ is balanced for $1 \leq u \leq s$ and their union topology $\mathcal{G}_{\mathcal{C}}$ contains a spanning tree. Besides, the Markov chain
$m_k$ is homogeneous and ergodic with a station distribution $\pi_u=\lim_{k\rightarrow\infty}\mathbb{P}(m_k=u)$ for all $u\in \mathcal{S}$, where $\sum_{u=1}^s\pi_u=1$.
\end{assumption}
\vspace{-5pt}
\begin{rema}\label{RAM}
Define $p_{vu}=\mathbb{P}(m_{k+1}=v \mid m_k=u)$, $q_{ij,k}=\mathbb{P}((i,j)\in \mathcal{E}_k)$ and $P_{u,k}=\mathbb{P}(\mathcal{G}_k=\mathcal{G}^{(u)})$. Then, it can be got $P_{u,k+1}=\sum_{v=1}^sP_{v,k}p_{vu}$. Define $G_{ij}=\{u\in {\mathcal{S}}|(i,j)\in \mathcal{E}^{(u)} \}$. Then, we can get $q_{ij,k}=\sum_{u\in G_{ij}}P_{u,k}$ for any $k\geq 1$. From Theorem 1.2 on \cite{S_Seneta2006}, there exists $\lambda_{q}\in (0,1)$ such that $q_{ij,k}=\sum_{u\in G_{ij}}\pi_u+O(\lambda_{q}^k)$. Moreover, if Assumption \ref{AG} hold, then it can be seen
\vspace{-5pt}
\begin{align}\label{pi_min}
\underline{\pi}\triangleq \min_{(i,j)\in\mathcal{E}_{\mathcal{C}}}\sum_{u\in G_{ij}}\pi_u>0.
\end{align}
\end{rema}
\vspace{-10pt}
\begin{assumption}\label{AR}
\textbf{(Conditional expectation type cooperative excitation condition)}
The regressor $\{\phi_{k,i},$ $i=1,\ldots,m\}$ satisfy $\|\phi_{k,i}\| \leq \bar{\phi}<\infty$ and there exists a positive integer $h$ and a positive number $\delta_{\phi}$ such that
\begin{gather}
\frac{1}{mh}\mathbb{E}\bigg[\sum^{m}_{i=1} \sum^{k+h-1}_{l=k}\phi_{l,i}\phi_{l,i}^T\bigg| \mathcal{F}_{k-1}\bigg]\geq\delta_{\phi}^2I_{n},  \forall k\geq 1.\nonumber
\end{gather}
\end{assumption}

\begin{rema}\label{RAR}
Assumption \ref{AR} is the conditional expectation type cooperative excitation condition, which is more general than cooperative excitation condition (i.e., $\mathbb{E}[\sum^{m}_{i=1}\phi_{k,i}\phi_{k,i}^T]>0$) in \cite{FCZ2022,KMR2012,LC2020}. Moreover, it includes the periodical full-rank condition \citep{WZY2003,WTZ2018,ZZK2023} and the deterministic persistent excitation condition \citep{GZ2013,WZZG2022} in the existing quantized identification work.
\end{rema}

\begin{assumption}\label{AP}
There is a known bounded convex compact set $\Omega\subset\mathbb{R}^n$ such that
the unknown parameter $\theta\in \Omega$. And denote $\bar{\theta}=\sup_{\eta\in\Omega}\|\eta\|.$
\end{assumption}

\begin{assumption}\label{AD}
For the observation noise sequence $\{d_{k,i}\}$, its conditional probability distribution function given $\mathcal{F}_{k-1}$ is denoted by $F_{k,i}(\cdot)$.
Besides, the communication noise $\omega_{k,ij}$ is mutually independent with respect to $i$ and $j$, and the conditional probability distribution function of  $\omega_{k,ij}$ given $\mathcal{F}_{k-1}$ is denoted by $G_{k,ij}(\cdot)$.
\end{assumption}

\begin{rema}\label{RAR}
The prior information condition on the unknown parameter (i.e., Assumption \ref{AP}) is a common assumption of quantized identification studies \citep{GZ2013,WZZG2022,WZZ2021,ZWZ2021,ZZG2022}, primarily used to ensure the boundedness of the estimate.  In contrast to previous works \citep{GZ2013,KMR2012,WZZG2022,WZZ2021,FCZ2022,You2015}, Assumption \ref{AD} no longer requires the noise to be independently and identically distributed.
\end{rema}

The goal of this paper is to develop a distributed estimation algorithm and a suitable quantized communication mechanism to estimate the unknown parameter based on quantized measurement and quantized communication under Markov-switching digraphs.

\section{Algorithm Design}\label{sec:EFQDIA}

For the quantized system (\ref{M}) with quantized communication (\ref{est_c}), handling the dual quantized information is the main challenge in distributed estimation. For quantized measurements, we can adopt the approach in \cite{WZZ2021,FCZ2022}, using noise’s statistical properties to recover unknown and time-invariant parameters. In quantized communication, the existing researches often employ probabilistic quantizer-based communication \citep{KMR2012,ZSX2015,ZCGXLJ2018}, which limit quantized error with zero mean, preserving information in expectation though requiring bit count proportional to data size. In contrast, this paper aims to achieve 1-bit quantized communication, resulting in unbounded quantized error. The core challenge, then, is how to recover high-dimensional, time-varying information about neighboring nodes using only 1-bit quantized communication.

This paper addresses this challenge by designing an estimator based on quantized communication, leveraging stochastic approximation method to estimate neighboring estimates. To accurately reconstruct the original neighbor estimates from quantized communication, the communication rule must satisfy an excitation condition. For simplicity, we employ a linear communication rule that meets the persistent excitation condition \citep{GZ2013,WZZG2022}. Using these estimated values in place of the transmitted neighboring estimates, a consensus strategy is developed to construct a distributed identification algorithm. The resulting estimation-fusion type quantized distributed identification (EFTQDI) algorithm  is presented as follows.
\vspace{-5pt}
\begin{algorithm}[htb]
\caption{EFTQDI algorithm}
\label{Algorithm}
For any given sensor $i\in\{1,2,\ldots,m\}$, begin with initial estimates  $\hat{\theta}_{0,i}\in\mathbb{R}^n$ and ${\theta}_{0,i}\in\mathbb{R}^{n}$. The algorithm is recursively defined as follows:
\begin{algorithmic}[1]
\State Design of communication rule (design the compression recoding function $Q(\cdot)$ as a linear rule):
\begin{equation}
\begin{aligned}
&Q(\theta_{k-1,j})=\psi_{k}^T\theta_{k-1,j},  j=1,\ldots,m, \nonumber
\end{aligned}
\end{equation}
where the encoding rule $\{\psi_{k}\}$ satisfy $\|\psi_{k}\| \leq \bar{\psi}<\infty$ and there exists a positive constant $\delta_{\psi}$ such that $\frac{1}{h}\sum^{k+h-1}_{l=k}\psi_{l}\psi_{l}^T \geq\delta_{\psi}^2I_{n}$ for all $k\geq 1$. Then, the quantized communication mechanism (\ref{est_c}) is
\begin{equation}\label{est_z}
\begin{aligned}
&z_{k,ij}=I_{\{\psi_{k}^T\theta_{k-1,j}+\omega_{k,ij}\leq C_{ij}\}}, \forall j\in\mathcal{N}_{i,k}.
\end{aligned}
\end{equation}
\State Estimation (establish the estimate of neighboring estimate based on quantized communication (\ref{est_z})):
\begin{align}\label{AE}
\hat{\theta}_{k,ij}=
\left\{
   \begin{array}{l}
    \Pi_{\Omega}\big\{\hat{\theta}_{k-1,ij}
+\gamma b_k\psi_k(\hat {G}_{k,ij}-z_{k,ij})\big\}\\
  \hbox{~~~~~~~~~~~~~~if } ~ j\in\mathcal{N}_{i,k}; \\
     \hat{\theta}_{k-1,ij}, \hbox{~~~if } ~ j\in \mathcal{N}_{i}^{\mathcal{C}}\setminus \mathcal{N}_{i,k},
   \end{array}
\right.
\end{align}
where the projection $\Pi_{\Omega}(\cdot)$ is defined as $\Pi_{\Omega}(x)= \arg\min_{z\in\Omega}\|x-z\|$ for all $x\in\mathbb{R}^n$,  $b_k>0$ is a step size, $\gamma>0$ is its step coefficient to be designed,  $\hat{G}_{k,ij}={G}_{k,ij}(C_{ij}-\psi_k^T \hat{\theta}_{k-1,ij})$,  and  $\mathcal{N}_{i}^{\mathcal{C}}$ is the neighbor set of the sensor $i$ in the union topology $\mathcal{G}_{\mathcal{C}}$.
\label{code:fram:extract}

\State Fusion (generate ${\theta}_{k,i}$  based on the estimates of neighboring estimate $\hat{\theta}_{k,j}^{i}$ and a consensus strategy):
\begin{align}\label{AF}
{\theta}_{k,i}=\Pi_{\Omega}\bigg\{&{\theta}_{k-1,i}+\beta b_k\bigg(\phi_{k,i} (\hat{F}_{k,i}-s_{k,i})\nonumber\\
&+\sum_{j\in \mathcal{N}_{i,k}}a_{ij,k} (\hat{\theta}_{k-1,ij}-{\theta}_{k-1,i})\bigg)\bigg\},
\end{align}
where $\beta>0$ is a step coefficient to be designed and $\hat{F}_{k,i}=F_{k,i}(C_i-\phi^T_{k,i}\theta_{k-1,i})$.
\label{code:fram:trainbase}
\end{algorithmic}
\end{algorithm}

\begin{rema}\label{RSALS}
This paper presents a quantized distributed algorithm based on a stochastic approximation approach rather than least squares or stochastic gradient methods. While the latter can relax regressor conditions and speed up convergence in distributed settings, they rely on the extensive sharing of regressor information among individual sensors \citep{GL2022,XZG2021,WZZ2021}. In contrast, this setup need to achieve distributed identification with minimal information exchange via quantized communication. The stochastic approximation algorithm enables efficient operation in this setup by only requiring neighboring sensors to share estimates, without transmitting regressor information \citep{ZZ2012,LC2020}. Consequently, this approach is well-suited to information-limited environments.
\end{rema}

\begin{rema}\label{RAC}
The compression coding coefficients $\psi_k$ in the linear communication encoding rule is designed to satisfy the persistent excitation condition \citep{GZ2013,WZZG2022}, which is made to ensure that the original vector information can be recovered through the compressed quantized communication information. A simple example meeting this encoding-rule is that periodically select $\psi_{k}$ from the set $\{e_1, e_2,\ldots, e_n\}$, where $e_i=[0,\ldots,0,{1},0,\ldots,0]\in\mathbb{R}^n$ is the standard unit basis vectors of Euclidean space $\mathbb{R}^n$. In fact, the condition of the compression encoding coefficients $\psi_k$ can be generalized to a conditional expectation-based persistent excitation condition. Specifically, there exists a positive integer $h_{\psi}$ and a positive number $\delta_{\psi}$ such that
$\frac{1}{h_{\psi}}\mathbb{E}\big[\sum^{k+h_{\psi}-1}_{l=k}\psi_{l}\psi_{l}^T\big| \mathcal{F}_{k-1}\big]\leq\delta_{\psi}^2I_{n}$, where $h_{\psi}$ could be different with $h$ in Assumption \ref{AR}.
\end{rema}

\begin{rema}\label{RLk}
The adjacency matrix $\mathcal{A}_k=\{a_{ij,k}\}$ of the communication topology $\mathcal{G}_k$ has the following properties under Assumption \ref{AG}: there exists $\lambda_{a}\in (0,1)$ such that $\mathbb{E}a_{ij,k}=\sum_{u=1}^s\pi_ua_{ij}^{(u)}+O(\lambda_{a}^k)$, which can be derived by Remark \ref{RAM} and Theorem 1.2 on \cite{S_Seneta2006}. Therefore, its adjacency matrix satisfies $\mathbb{E}\mathcal{A}_k=\sum_{u=1}^s\pi_u\mathcal{A}^{(u)}+O(\lambda_{a}^k)$ and its Laplacian matrix $\mathcal{L}_k$ satisfies $\mathbb{E}\mathcal{L}_k=\sum_{u=1}^s\pi_u\mathcal{L}^{(u)}+O(\lambda_{a}^k)$.
\end{rema}

Next, we give the design requirement about the step size $b_k$ in the EFTQDI algorithm as follows.
\begin{assumption}\label{Ab}
The step size $b_k$ satisfies:  $\sum_{k=0}^{\infty} b_k=\infty$, $\lim _{k \rightarrow \infty} b_k=0$ and $b_k=O\( b_{k+1}\)$.
\end{assumption}
\begin{rema}\label{RAb}
The conditions of $\sum_{k=0}^{\infty} b_k=\infty$ and $\lim _{k \rightarrow \infty} b_k=0$ are common requirements for stochastic approximation type algorithms \citep{RM1951,LC2020,ZZ2012}. Besides,  $b_k=O\( b_{k+1}\)$ guarantees that the step sizes of the algorithm do not vary significantly over a finite number of steps, which is useful for handling the conditional expectation-type cooperative excitation condition. Moreover, Assumption \ref{Ab} shows $1<\sup_{|p-q|<h}{b_{p}}/{b_{q}}<\infty$.
\end{rema}

\section{Main results}\label{sec:conver}
To clearly present the findings of this paper, this section outlines the main results, including the mean square convergence and convergence rate of the proposed algorithm. Proofs of these results will be provided collectively in the following section. Specifically, Section 4.1 introduces the common notations used in the theoretical analysis, while Section 4.2 presents the key results.


\subsection{Notation}\label{sec:nota}
For convenience of analysis, we consider the agent $i$ within the union graph $\mathcal{G}_{\mathcal{C}}$, denote $N_i$ as the size of its neighbor set $\mathcal{N}_{i}^{\mathcal{C}}$ and $N=\sum_{i=1}^{m}N_{i}$.
Denote ${a}_{ij}^{\mathcal{C}}=\sum_{u=1}^s\pi_ua_{ij}^{(u)}\in[0,1)$ in its adjacency matrix
$\mathcal{A}_{\mathcal{C}}$. Accordingly, its Laplacian matrix is given by ${\mathcal{L}}_{\mathcal{C}}=\sum_{u=1}^s\pi_u\mathcal{L}^{(u)}$.

Let $\varepsilon_{k,ij}=\hat{\theta}_{k,ij} -\theta_{k,j}$ as the estimation error of neighboring estimate. Putting $\varepsilon_{k,ij},j\in \mathcal{N}_i^{\mathcal{C}}$ in a given order yields the error vector $\varepsilon_k$. Without loss of generality, let
\begin{align}\label{NE}
 \varepsilon_k=(&\varepsilon_{k,1r_1},\ldots, \varepsilon_{k,1r_{N_1}},\varepsilon_{k,2r_{N_1+1}}, \ldots,\varepsilon_{k,2r_{N_1+N_2}},\ldots,\nonumber\\
 &\varepsilon_{k,mr_{N_1+\cdots+N_{m-1}+1}}, \ldots,\varepsilon_{k,mr_{N_1+\cdots+N_{m}}})^T,
\end{align}
where $r_1,\ldots,r_{N_1}\in\mathcal{N}_1^{\mathcal{C}},r_{N_1+1}, \ldots,r_{N_1+N_2}\in\mathcal{N}_2^{\mathcal{C}},\ldots,$ $r_{N_1+\cdots+N_{m-1}+1},\ldots,r_{N_1+\cdots+N_m} \in\mathcal{N}_m^{\mathcal{C}}$.

Then, we construct three matrices to establish the corresponding relation of the fusion estimate and its estimates.

$R_k$ is designed to select each neighbor of each sensor node at time $k$ in the union graph $\mathcal{G}_{\mathcal{C}}$.
Define $R_{k}=\diag\{\bar{r}_{k,1},\ldots, \bar{r}_{k,N}\}\in \mathbb{R}^{N \times N}$. Let $r_{l}$ in vector $\varepsilon_k$ represents the neighbor $j$ of agent $i$, i.e., $\varepsilon_{k,ir_{l}}=\varepsilon_{k,ij}$,
where $r_{l}\in \mathcal{N}_{i}^{\mathcal{C}}$ and $l\in \{N_{1}+\cdots+N_{i-1}+1,\ldots, N_{1}+\cdots+N_{i}\}$.
Then, $\bar{r}_{k,l}=\sign\{a_{ij,k}\}=\sign\{a_{ir_{l},k}\}$ for $l\in\{1,\ldots,N \}$, where $\sign\{x\}$ is the sign of scalar $x$ and $a_{ij,k}$ is the element of the adjacency matrix $\mathcal{A}_k$.

$P_k$ is designed to select the neighbor set of each node at time $k$.
Define $P_k=[p_{k,1},\ldots, p_{k,m}]^{T}\in \mathbb{R}^{m \times N},$ where
$p_{k,i}=[0,\ldots,0,\underbrace{a_{ir_{N_1+\cdots+N_{i-1}+1},k},\ldots, a_{ir_{N_1+\cdots+N_{i}},k},}_{N_1+\cdots+ N_{i-1}+1\text{th to } N_1+\cdots+N_{i}\text{th position}}$ $0,\ldots,0]^T\in \mathbb{R}^{N}$ for $i=1,\ldots,m$.

$Q$ is designed to select the true neighboring estimate of each node that correlates with its estimate.
Define $Q=[q_{1,r_1}, \ldots, q_{1,r_{N_1}}, \ldots, q_{m,r_{N_1+\cdots+N_{m-1}+1}}, \ldots, q_{m,r_{N_1+\cdots+N_{m}}}]^{T}$ $\in \mathbb{R}^{N \times m}$,
where $q_{i,j}=[0,\ldots,0,\underbrace{1}_{j\text{th position}},0,\ldots,0]^T\in \mathbb{R}^{m}$ for  $(j,i)\in\mathcal{E}^{\mathcal{C}}$.

Next, we introduce the following notations.
\vspace{-7pt}
\begin{align}
&S_k=\col\{s_{k,1},\ldots,s_{k,m}\}\in\mathbb{R}^{m} \nonumber\\
&\Phi_k=\diag\{\phi_{k,1},\ldots,\phi_{k,m}\}\in\mathbb{R}^{mn\times m}\nonumber\\
&\Psi_k=\col\{\psi_{k},\ldots,\psi_{k}\}\in\mathbb{R}^{nN\times N} \nonumber\\
&{\Theta}_k=\col\{\theta_{k,1},\ldots,\theta_{k,m}\}\in\mathbb{R}^{mn\times 1} \nonumber\\
&\tilde{\Theta}_k=\col\{\tilde{\theta}_{k,1},\ldots,\tilde{\theta}_{k,m}\}\in\mathbb{R}^{mn\times 1}, \text{ where } \tilde{\theta}_{k,i}=\theta_{k,i}-\theta. \nonumber\\
&\hat{F}_k=\col\{\hat{F}_{k,1},\ldots,\hat{F}_{k,m}\}\in\mathbb{R}^{m\times 1} \nonumber\\
&\hat{\Theta}_k=\col\{\hat{\theta}_{k,1r_1},\ldots,\hat{\theta}_{k,1r_{N_1}},\ldots, \hat{\theta}_{k,mr_{N_1+\cdots+N_{m-1}+1}},\ldots, \nonumber\\ &\qquad\quad\quad~\hat{\theta}_{k,mr_{N_1+\cdots+N_{m}}}\}\in\mathbb{R}^{nN\times 1} \nonumber\\
&{Z}_k=\col\{z_{k,1r_1},\ldots, z_{k,1r_{N_1}}, \ldots, z_{k,mr_{N_1+\cdots+N_{m-1}+1}}, \ldots, \nonumber\\ &\qquad\quad\quad~z_{k,mr_{N_1+\cdots+N_{m}}}\}\in\mathbb{R}^{N} \nonumber\\
&\hat{G}_k=\col\{\hat{G}_{k,1r_1},\ldots,\hat{G}_{k,1r_{N_1}},\ldots,\hat{G}_{k,mr_{N_1+\cdots+N_{m-1}+1}}, \nonumber\\
&\qquad\quad\quad~ \ldots,\hat{G}_{k,mr_{N_1+\cdots+N_{m}}}\}\in\mathbb{R}^{N} \nonumber
\end{align}

\vspace{-5pt}
where $\col\{\cdot,\ldots,\cdot\}$ denotes the vector stacked by the specified vectors, $\diag\{\cdot,\ldots,\cdot\}$ represents a block matrix arranged diagonally with the given vectors or matrices.
For any $(j,i)\in \mathcal{E}_{\mathcal{C}}\setminus \mathcal{E}_{k}$, we set $z_{k,ij}=\hat{G}_{k,ij}=-1$ for all  without loss of generality. Actually, $z_{k,ij}$ and $\hat{G}_{k,ij}$ for $(j,i)\in \mathcal{E}_{\mathcal{C}}\setminus \mathcal{E}_{k}$ can be assigned any constant value without affecting the algorithm, as shown in (\ref{AE}), which indicates that the proposed algorithm does not depend on $z_{k,ij}$ and $\hat{G}_{k,ij}$ when $(j,i)\in \mathcal{E}_{\mathcal{C}}\setminus \mathcal{E}_{k}$.

Based on the above notations, the EFTQDI algorithm (\ref{AE})-(\ref{AF}) can be written as
\vspace{-5pt}
\begin{align}\label{EFQDIA}
\left\{
\begin{aligned}
\hat{\Theta}_{k}=&\mathbf{\Pi}\left\{\hat{\Theta}_{k-1}+\gamma b_k{R}_k\otimes I_n\Psi_k(\hat{G}_{k}-Z_{k})\right\},\\
{\Theta}_{k}=&\mathbf{\Pi}\big\{{\Theta}_{k-1}+\beta b_k\Phi_k(\hat{F}_{k}-S_{k})-\beta b_k\mathcal{L}_k\\
&\quad\otimes I_n{\Theta}_{k-1}+\beta b_kP_k\otimes I_n\varepsilon_{k-1}\big\}.
\end{aligned}
\right.
\end{align}
where $\mathbf{\Pi}\left\{\cdot\right\}$ is defined as $\mathbf{\Pi}\left\{\zeta\right\}\triangleq\left({\Pi}\left\{\zeta_1\right\},\ldots,
{\Pi}\left\{\zeta_n\right\}\right)^T$ for $\zeta=\col\{\zeta_1,\ldots,\zeta_n\}$, and $\varepsilon_ {k}=\hat{\Theta}_{k}-Q\otimes I_n{\Theta}_{k}$.

\subsection{Convergence properties}
It is worth noticing that the fusion estimate $\theta_{k,i}$ is related to the estimates of neighboring estimates $\hat{\theta}_{k-1,ij}$, which complicates the convergence analysis of both estimates. To deal with the coupling between these two estimates, we introduce two Lyapunov functions for the estimation error as follows,
\vspace{-5pt}
 \begin{align}\label{UkVk}
U_k= \mathbb{E}[\varepsilon^{T}_k\varepsilon_k],\quad 
V_k=\mathbb{E}[\tilde{\Theta}^{T}_k\tilde{\Theta}_k]. 
\end{align}
Then, we will establish the coupling relationship between these two estimates by jointly analyzing these two Lyapunov functions. First, we introduce the following conditions regarding the observation noise $d_{k,i}$ and communication noise $\omega_{k,ij}$.

\begin{assumption}\label{ADD}
The corresponding condition density functions $f_{k,i}(\cdot)$ and $g_{k,ij}(\cdot)$ given $\mathcal{F}_{k-1}$ of the observation noise $d_{k,i}$ and the communication noise $\omega_{k,ij}$ satisfy
\begin{align}
&\underline{f}=\min_{i}\inf_{k\geq 1}\inf_{|C_i-x|\leq\bar{\phi}\bar{\theta}} f_{k,i}(x)>0, \nonumber\\ &\bar{f}=\max_{i}\sup_{k\geq 1}\sup_{|C_i-x|\leq\bar{\phi}\bar{\theta}} f_{k,i}(x)<\infty, \nonumber\\
&\underline{g}=\min_{(j,i)\in\mathcal{E}_{\mathcal{C}}}\inf_{k\geq 1} \inf_{|C_{ij}-x|\leq\bar{\psi}\bar{\theta}} g_{k,ij}(x)>0.\nonumber
\end{align}
\end{assumption}

Then, we present the following lemma to demonstrate the relationship between the Lyapunov function $U_k$ for the fusion estimation error and the Lyapunov function $V_k$ for the neighboring estimation error.

\begin{lemma}\label{lemmaUk}
If Assumptions \ref{AG}, \ref{AP}-\ref{ADD} hold, then the Lyapunov function $U_k$ for fusion estimation error satisfies
\vspace{-20pt}
\begin{small}
\begin{align}
U_k\leq &\(1-\(\frac{2\gamma \underline{g}\underline{\pi}\delta_{\psi}^2}{c_h}-\beta\(\frac{\bar{f} \bar{\phi}^2}{\alpha_1}+2\bar{N}+\frac{\bar{N}}{\alpha_2}\)\)\sum_{l=k-h}^{k-1}b_{l+1}\)U_{k-h}\nonumber\\
&+\beta(\alpha_1\bar{f}\bar{N} \bar{\phi}^2+\alpha_2\lambda_m^2) \sum_{l=k-h}^{k-1}b_{l+1}V_{k-h} +O\(b_k^2\),\nonumber
\end{align}
\end{small}
for all $\alpha_1,\alpha_2 \in(0,\infty)$, where $\bar{N}=\max_{i=1,\ldots,m}N_{i}$ is the biggest size of the neighbor set in the union topology $\mathcal{G}_{\mathcal{C}}$,  $\lambda_m\triangleq\max_{u=1,\ldots,s}\|\mathcal{L}^{(u)}\|$, $c_{h}\triangleq\sup_{|p-q|<h}\frac{b_{p}}{b_{q}}$, $\underline{\pi}$, $\bar{\phi}$, $\underline{g}$, $\bar{f}$, $\delta_{\psi}$ are defined in (\ref{pi_min}), Assumptions \ref{AR}, \ref{ADD} and Algorithm \ref{AG}.
\end{lemma}
\begin{rema}\label{RAU}
Lemma \ref{lemmaUk} provides the iterative relationship between $U_k$, $U_{k-h}$ and $ V_{k-h}$, showing that with appropriate step coefficient design, $ U_k$ can be reduced relative to $U_{k-h}$, while still being influenced by the coupling term $V_{k-h}$.
The reason we establish an iterative relationship between $U_k$ and $U_{k-h}$ rather than $U_k$ and $U_{k-1}$ is that, in the latter case (i.e., Eq.(\ref{varepsilon})), the encoding rule $\psi_{k}$ does not satisfy the persistent excitation condition under one iteration. This prevents us from establishing a compression relationship between $U_k$ and $U_{k-1}$, resulting in divergence due to the perturbation term $V_{k-1}$ and making convergence analysis of $U_k$ infeasible.
However, when analyzing the relationship between $U_k$ and $U_{k-h}$, the encoding rule $\psi_{k}$, under the persistent excitation condition, provides a compression coefficient (i.e., Eq. (\ref{appA11})), enabling a compression relationship between $U_k$ and $U_{k-h}$ with proper step coefficient design.
\end{rema}

\begin{rema}\label{RAU2}
This lemma also supports the view expressed in Remark \ref{RAC}, which states that the encoding rule $\psi_{k}$ can be extended to satisfy $\|\psi_{k}\| \leq \bar{\psi}<\infty$ and
$\frac{1}{h_{\psi}}\mathbb{E}\big[\sum^{k+h_{\psi}-1}_{l=k}\psi_{l}\psi_{l}^T\big| \mathcal{F}_{k-1}\big]\leq\delta_{\psi}^2I_{n\times n}$ for all $k\geq 1$. Its proof is similar as  (\ref{appA10})-(\ref{appA11}) and is omitted. Under this condition, Lemma \ref{lemmaUk} establishes the iterative relationship between $U_k$, $U_{k-h_{\psi}}$ and $ V_{k-h_{\psi}}$.
\end{rema}

Next, we present the following lemma to illustrate how the iteration of the Lyapunov function $V_k$ is influenced by  the Lyapunov function $U_k$ for fusion estimation error.
\vspace{-10pt}
\begin{lemma} \label{lemmaV_k}
If Assumptions \ref{AG}-\ref{AR} and \ref{AD}-\ref{ADD} hold, then the Lyapunov function $V_k$ for the neighboring estimation error satisfies that
\begin{align}
  V_k\leq& \(1-\beta\(\frac{2\sigma}{c_h}-\frac{1}{\alpha_3}\)\sum_{l=k-h}^{k-1}b_{l+1}\)V_{k-h}\nonumber\\
  &+\alpha_3\beta \bar{N}\sum_{l=k-h}^{k-1}b_{l+1}U_{k-h}+O\(b_{k}^2\),\nonumber
\end{align}
for all $\alpha_3\in(0,\infty)$, where $\sigma=\frac{h\underline{f}\hat{\lambda}_2\delta_{\phi}^2}{2\underline{f}\bar{\phi}^2
+h\hat{\lambda}_2}$, $\hat{\lambda}_2$ is the smallest nonzero eigenvalue of the Laplacian matrix
$\hat{\mathcal{L}}_{\mathcal{C}}$ of the mirror graph for the union graph $\mathcal{G}_{\mathcal{C}}$, $c_h$, $\delta_{\phi}$, $\bar{\phi}$ and $\underline{f}$ are defined in Lemma \ref{lemmaUk},Assumptions \ref{AR} and \ref{ADD}.
\end{lemma}
\vspace{-5pt}
\begin{rema}\label{RAV}
Lemma \ref{lemmaV_k} presents the iterative relationship between $V_k$, $V_{k-h}$ and $ U_{k-h}$, showing that with appropriate step coefficient design, $V_k$ can be reduced relative to $V_{k-h}$, while still being perturbed by the coupling term $U_{k-h}$. Together with Remark \ref{RAU}, this implies that we cannot analyze the convergence of $U_k$ or $V_k$ independently; instead, we must jointly analyze both $U_k$ and $V_k$ with respect to their common coupling terms.
\end{rema}
\vspace{-5pt}
\begin{rema}\label{RAV2}
Similar with Remark \ref{RAU}, the conditional expectation-type cooperative excitation condition of the regressor $\phi_{k,i}$ guarantees the compression relationship between $V_k$ and $V_{k-h}$. Specifically, this condition, combined with the connectivity of the mirror graph of the union graph $\mathcal{G}_{\mathcal{C}}$, guarantees the compression term $2\beta\sigma\sum_{l=k-h}^{k-1}b_{l+1}V_{k-h}$. This implies that even without transmitting the regressor information, the collective effect of the sensors can still be achieved solely through the quantized information exchanged between sensors. In other words, while each sensor cannot estimate the global information based solely on its local measurements, the exchange of one-bit information between sensors allows each sensor to reconstruct the global information.
\end{rema}

Building on the above lemmas, the following theorem demonstrates the mean square convergence of the fusion estimate of the unknown parameter given by (\ref{AF}) and the estimate of the neighboring estimate given by (\ref{AE}).
\vspace{-5pt}
\begin{theorem}\label{thm_c}
If Assumptions \ref{AG}-\ref{ADD} hold, then the proposed EFTQDI algorithm is convergent in the mean square sense, i.e.,
 \begin{align}
\lim_{k\rightarrow\infty}\mathbb{E}\|\tilde{\Theta}_k\|^2=0 \text{ and } \lim_{k\rightarrow\infty}\mathbb{E}\|\varepsilon_k\|^2=0,\nonumber
 \end{align}
providing that $\gamma>\frac{\beta}{2\underline{g}\underline{\pi}\delta_{\psi}^2}
\big(2c_h\bar{N}+\sigma\(\bar{f}^2 \bar{\phi}^4+\lambda_{m}^2\)+\frac{8c_h^4\bar{N}^2}{3\sigma^3} \big)$,
where $\underline{\pi}$, $\sigma$, $\hat{\lambda}_2$, $\bar{N}$, $\lambda_{m}$, $\underline{g}$,  $\bar{f}$, $\delta_{\psi}$ are defined in Lemmas \ref{lemmaUk}-\ref{lemmaV_k}, respectively.
\end{theorem}

\begin{rema}\label{RTC}
To prove the mean-square convergence of the two estimates in the propose algorithm, it suffices to show that the constructed Lyapunov functions $U_k$ and $V_{k}$ are convergent. Since they are coupled, we jointly analyze their convergence properties of the two inequalities in Lemmas \ref{lemmaUk} and \ref{lemmaV_k}.
For mean-square convergence, the step-size coefficient $\gamma$ of the estimate of neighboring estimate must be in a certain proportional relationship with $\beta$, the step-size for the fusion estimate. This is because the estimate for neighboring estimate $\hat{\theta}_{k,ij}$ essentially tracks the fusion estimate ${\theta}_{k,j}$  generated by its neighbors. When the step-size coefficient $\beta$ of the fusion estimate increases, the step-size coefficient $\gamma$ of the estimate for neighboring estimate $\hat{\theta}_{k,ij}$ must also be adjusted accordingly to effectively track the slow time-varying fusion estimate ${\theta}_{k,j}$.
\end{rema}

Finally, the following theorem demonstrates the mean square convergence rate for the case where the step-size is of the polynomial form $b_k=\frac{1}{k^p},p\in(1/2,1]$.
\vspace{-5pt}
\begin{theorem}\label{thm_cr}
Under the conditions of Theorem \ref{thm_c}, the proposed EFTQDI algorithm has the following mean square convergence rate:
\begin{itemize}
\item For the case of the step-size as $b_k=\frac{1}{k^p}\( \frac{1}{2}<p<1\)$,  $\mathbb{E}\|\tilde{\Theta}_k\|^2= O\left(\frac{1}{k^p}\right)$ and $\mathbb{E}\|\varepsilon_k\|^2= O\left(\frac{1}{k^p}\right)$.
\item For the case of the step-size as $b_k=\frac{1}{k}$, it can been seen that  $\mathbb{E}\|\tilde{\Theta}_k\|^2 = O\left(\frac{1}{k}\right)$ and $\mathbb{E}\|\varepsilon_k\|^2= O\left(\frac{1}{k}\right)$ providing that $\beta>\frac{2c_h}{3\sigma}$ and $\gamma>\frac{\beta}{2\underline{g}\underline{\pi}\delta_{\psi}^2}
\big(2c_h\bar{N}+\sigma\(\bar{f}^2 \bar{\phi}^4+\lambda_{m}^2\) +\frac{8\beta c_h^4\bar{N}^2}{\sigma^2({3}\beta\sigma-2c_h)}+
\frac{c_h}{\beta} \big)$.
\end{itemize}
\end{theorem}

\begin{rema}\label{RTC}
It is important to noting that when the step size is $b_k=\frac{1}{k}$, achieving the convergence rate at the same order as itself requires additional constraints on the step-size coefficient $\gamma$ and $\beta$, due to the unique properties of the harmonic series $\sum\frac{1}{k}$. Specifically, the harmonic series is highly sensitive, and its coefficient $\tau$ can affect the order of the product $\prod_{k=1}^l\left(1-\frac{\tau}{k}\right)$ while the product $\prod_{k=1}^l\left(1-\frac{\tau}{k^p}\right)$ for $\frac{1}{2}<p<1$ is not similarly influenced.
Specifically, to guarantee the convergence rate as $O(b_k)$ for the case of $b_k=\frac{1}{k}$, the step-size coefficient $\beta$ of the fusion estimate is in inverse proportion to the connectivity $\hat{\lambda}_2$ of the mirror graph of the union graph $\mathcal{G}_{\mathcal{C}}$ and stochastic measurement noise coefficient $\underline{f}$. And the step-size coefficient $\gamma$ for the estimate of the neighboring estimate is inversely proportional to communication uncertainties including Markov-switching rate $\underline{\pi}$ and stochastic communication noise coefficient $\underline{g}$.
\end{rema}

\vspace{-5pt}
\section{Proofs of the main results}\label{sec:proof}
In this section, some lemmas are collected and eatablished, which are frequently used in the analysis of convergence and convergence rate.
\begin{lemma}\citep{CM1987}\label{lem:proj}
For the bounded convex set $\Omega$, the projection is defined as $\Pi(x)= \arg\min_{z\in\Omega}||x-z||$ for all $x\in\mathbb{R}^n$.  Then, for all $x\in\mathbb{R}$ and $x^*\in \Omega$, it holds $\left\|\Pi(x)-x^*\right\|\leq \left\|x-x^*\right\|$.
\end{lemma}

\begin{lemma}\citep{RB2005}\label{lem:hatL}
Let $\mathcal{G}= (\mathcal{V}, \mathcal{E},\mathcal{A})$ be a directed graph with the Laplacian matrix $\mathcal{L}$. Then, $\frac{\mathcal{L}+ \mathcal{L}^T}{2}$ is the Laplacian matrix of its mirror graph $\hat{\mathcal{G}}$  if and only if $\mathcal{G}$ is a balanced graph.
\end{lemma}

\begin{lemma} \label{lemmaAM}\citep{XG2018S}
Let $A_k=\diag\{A^1_k,\ldots,$ $A^m_k\}$, where $A^i_k\in\mathbb{R}^{n\times n}$ with $0\leq A^i_k\leq bI_{n}$, $k\geq 0,i=1,\ldots,m$, and $b>0$ is a constant.
If $\mathcal{L}$ is the Laplacian matrix of the undirect and connect topology, then for all $k\geq 0$ and $h\geq 1$,
\vspace{-20pt}
\begin{small}
$$\lambda_{\min}\left(\sum^{k+{h}-1}_{l=k}\left[A_l+
\nu\mathcal{L}\right]\right)\geq \frac{\nu \lambda_2}{m(2b+\nu \lambda_2)}\lambda_{\min}\left(\sum^m_{i=1} \sum^{k+{h}-1}_{l=k}A^i_l\right),$$
\end{small}
where $\nu>0$, and $\lambda_2$ is the smallest nonzero eigenvalue of the Laplacian matrix $\mathcal{L}$.
\end{lemma}

\vspace{-5pt}
\begin{lemma}\citep{S_Polyak1987}\label{lem:xl}
Let $\{p_k\}$, $\{q_k\}$ and $\{\alpha_k\}$ be real sequences satisfying $p_{k+1} \leq(1-q_k)p_k+\alpha_k$, where $0<q_k\leq 1$, $\sum_{k=0}^{\infty} q_k=\infty$, $\alpha_k\geq 0$ and
$\lim_{k \rightarrow \infty}\frac{\alpha_k}{q_k}=0$.
Then,  $\limsup_{k \rightarrow \infty} p_k \leq 0$.
Particularly, if $p_k \geq 0$, then $\lim _{k \rightarrow \infty} p_k=0$.
\end{lemma}

\vspace{-5pt}
\begin{lemma}\citep{WKZ2024} \label{LemmaA23_WKZ2024}
For $0<b \leq 1, a>0, k_0 \geq 0$ and sufficiently large $l$, we have
\vspace{-25pt}
\begin{small}
$$
\begin{aligned}
& \prod_{i=l}^k\left(1-\frac{a}{\left(i+k_0\right)^b}\right)
\leq
\begin{cases}
\left(\frac{l+k_0}{k+k_0}\right)^a, & b=1; \\
e^{ \frac{a}{1-b}((l+k_0)^{1-b}-(k+k_0+1)^{1-b})},
 & b\in(0,1);
 \end{cases}\nonumber\\
&\sum_{l=1}^k\prod_{i=l}^k\left(1-\frac{a}{\left(i+k_0\right)^b}\right)
O\(\frac{1}{\left(l+k_0\right)^{2b}}\) =O\(\frac{1}{k^{b}}\), b\in(0,1).
\end{aligned}
$$
\end{small}
\end{lemma}

\vspace{-5pt}
\begin{lemma}\citep{ZWZ2021} \label{Lemma1_ZWZ2021}
For any given positive integer $l$ and $a, b\in \mathbb{R}$, the following results hold
\vspace{-25pt}
\begin{small}
\begin{align}
&\sum_{l=1}^{k}\prod_{i=l+1}^k\left(1-\frac{a}{i}\right)
\frac{1}{l^{1+b}}=
\frac{1}{l^{1+b}}=
\left\{
\begin{aligned}
&O\left(\frac{1}{k^a}\right),a<b,\\
&O\left(\frac{\ln k}{k^a}\right),a=b,\\
&O\left(\frac{1}{k^b}\right),a>b.
\end{aligned}
\right.\nonumber
\end{align}
\end{small}
\end{lemma}

\begin{lemma}\label{LemmaRPQ}
Under Assumption \ref{AG}, $R_k$, $P_k$ and $Q$ defined in Section \ref{sec:nota} have the following properties: \\
i) $\mathbb{E}R_k\geq \underline{\pi}I_{N}+O(\lambda_{q}^k)$;
ii) $\|P_k\|\leq \sqrt{\bar{N}}$;
iii) $\|Q\|\leq \sqrt{\bar{N}}$.
\end{lemma}
{\bf Proof}.
From the definition of $R_k$ and (\ref{pi_min}), we have
\vspace{-5pt}
\begin{align}
\mathbb{E}\bar{r}_{k,l}=&\mathbb{P}(r_l\in\mathcal{N}_{i,k})
=\mathbb{P}((i,r_l)\in\mathcal{E}_k)\nonumber\\
=&\sum_{u\in G_{ir_l}}\pi_u+O(\lambda_{q}^k)
\geq \underline{\pi}+O(\lambda_{q}^k),\nonumber
\end{align}
\vspace{-5pt}
for any $l=N_1+\cdots+N_{i-1}+1,\ldots,N_1+\cdots+N_{i}$. Thus,
\begin{align}
\mathbb{E}R_k=\diag\{&\mathbb{E}\bar{r}_{k,1},\ldots,\mathbb{E}\bar{r}_{k,N_1},\ldots,\mathbb{E}\bar{r}_{k,N_1+\cdots+N_{m-1}+1}, \nonumber\\
&\ldots,\mathbb{E}\bar{r}_{k,N_1+\cdots+N_m}\}\geq \underline{\pi}I_{N}+O(\lambda_{q}^k).\nonumber
\end{align}
From $a_{ij,k}\in[0,1)$, the definition of $P_k$ and $Q$, we have
\begin{align}
P_kP_k^T&=\diag\{p_{k,1}^{T}p_{k,1},\ldots, p_{k,m}^{T}p_{k,m}\}\nonumber\\
&=\diag\bigg\{\sum_{l=1}^{N_1}a_{1r_l,k}^2,\ldots, \sum_{l=N_1+\cdots+N_{m-1}+1}^{N_1+\cdots+N_{m}}a_{1r_l,k}^2\bigg\}\nonumber\\
&\leq\diag\{N_1,\ldots, N_m\}\leq \bar{N}I_m,\nonumber\\
Q^TQ=&\sum_{i=1}^{m}\sum_{j\in \mathcal{N}_{i}^{\mathcal{C}}} q_{ij}q_{ij}^{T}
=\diag\{N_1,\ldots, N_m\}
\leq \bar{N}I_m.\nonumber
\end{align}
\vspace{-5pt}
Then, we have
\vspace{-5pt}
\begin{align}
&\|P_k\|=\sqrt{\lambda_{\max}(P_k^TP_k)}=\sqrt{\lambda_{\max}(P_kP_k^T)}\leq \sqrt{\bar{N}}, \nonumber\\
&\|Q\|=\sqrt{\lambda_{\max}(Q^TQ)}\leq\sqrt{\bar{N}}.\nonumber
\end{align}
This completes the proof \qed

\vspace{-5pt}
\begin{lemma}\label{LemmaUVE}
Under Assumptions \ref{AR}, \ref{AP} and \ref{Ab}, $U_k$, $V_k$ and $\varepsilon_k$ satisfies: \\
i) $\left\|U_{k+l}-U_{k}\right\|=O(b_{k+l})$ for all $l\geq 0$;\\
ii) $\left\|V_{k+l}-V_{k}\right\|=O(b_{k+l})$ for all $l\geq 0$;\\
iii) $\left\|\varepsilon_{k+l}-\varepsilon_{k}\right\|=O(b_{k+l})$ for all $l\geq 0$.
\end{lemma}
\vspace{-5pt}
{\bf Proof}.
From Assumptions \ref{AR}-\ref{AP}, (\ref{AE}) and (\ref{AF}), we have
\vspace{-5pt}
\begin{align}
& \|\theta_{k,i}\|\leq \bar{\theta}, \quad \|\hat{\theta}_{k,ij}\|\leq \bar{\theta},  \label{appA1}\\
& \|\phi_{k,i}\|\leq \bar{\phi}, \quad \|\psi_{k}\|\leq \bar{\psi},  \label{appA2}
\end{align}
for all $i=1,\ldots,m$ and $j\in \mathcal{N}_{i}^{\mathcal{C}}$.
We now prove
\vspace{-5pt}
\begin{align}
&\left\|\tilde{\theta}_{k+l,i}-\tilde{\theta}_{k,i}\right\|\leq
\beta (\bar{\phi}+2N_i\bar{\theta})\sum_{p=k+1}^{k+l}b_p,\label{tildetheta}\\
&\left\|\varepsilon_{k+l,ij}-\varepsilon_{k,ij}\right\|\leq
(\gamma\bar{\psi}+\beta\bar{\phi}+2\beta N_i\bar{\theta})\sum_{p=k+1}^{k+l}b_p, \label{varepsilon}
\end{align}
for all $l\geq 0$, $i=1,\ldots,m$ and $j\in \mathcal{N}_{i}^{\mathcal{C}}$.
From Lemma \ref{lem:proj}, (\ref{AF}), (\ref{appA1}) and (\ref{appA2}), we have
\vspace{-5pt}
\begin{align}
&\left\|\tilde{\theta}_{k,i}-\tilde{\theta}_{k-1,i}\right\|\nonumber\\
=&\bigg\|\Pi_{\Omega}\bigg\{{\theta}_{k-1,i}+\beta b_k\big(\phi_{k,i} (F_{k,i}(C_i-\phi_{k,i}^T{\theta}_{k-1,i})-s_{k,i}) \nonumber\\
&+\sum_{j\in \mathcal{N}_{i,k}}a_{ij,k} (\hat{\theta}_{k-1,ij}-{\theta}_{k-1,i})\big)\bigg\}-{\theta}_{k-1,i}\bigg\|\nonumber\\
\leq&\bigg\|\beta b_k\big(\phi_{k,i} (F_{k,i}(C_i-\phi_{k,i}^T{\theta}_{k-1,i})-s_{k,i}) \nonumber\\
&+\sum_{j\in \mathcal{N}_{i,k}}a_{ij,k} (\hat{\theta}_{k-1,ij}-{\theta}_{k-1,i})\big)\bigg\|
\leq\beta (\bar{\phi}+2N_i\bar{\theta})b_k,\nonumber
\end{align}
\vspace{-5pt}
and hence
\vspace{-5pt}
\begin{align}
\left\|\tilde{\theta}_{k+l,i}-\tilde{\theta}_{k,i}\right\|
\leq&\sum_{p=k+1}^{k+l}\|\tilde{\theta}_{p,i}-\tilde{\theta}_{p-1,i}\|\nonumber\\
\leq&\beta (\bar{\phi}+2N_i\bar{\theta})\sum_{p=k+1}^{k+l}b_p.\nonumber
\end{align}
That is, (\ref{tildetheta}) hold. By Lemma \ref{lem:proj}, (\ref{AE}) and (\ref{appA1})-(\ref{tildetheta}), we have
\vspace{-5pt}
\begin{align}
&\left\|\varepsilon_{k,ij}-\varepsilon_{k-1,ij}\right\|\nonumber\\
=&\left\|\(\hat{\theta}_{k-1,ij}-{\theta}_{k,i}\) -\(\hat{\theta}_{k-1,ij}-{\theta}_{k-1,i}\)\right\|\nonumber\\
\leq&\left\|{\theta}_{k,i} -{\theta}_{k-1,i}\right\|\nonumber\\
\leq&\beta(\bar{\phi}+2 N_i\bar{\theta})b_k, \text{for } i\in \mathcal{N}_{i}^{\mathcal{C}}\setminus \mathcal{N}_{i,k}, \nonumber\\
&\left\|\varepsilon_{k,ij}-\varepsilon_{k-1,ij}\right\|\nonumber\\
=&\bigg\|\bigg(\Pi_{\Omega}\left\{\hat{\theta}_{k-1,ij}+\gamma b_k\psi_k\left(G_{k,ij}(C_{ij}-\psi_k^T \hat{\theta}_{k-1,ij})-z_{k,ij}\right)\right\}\nonumber\\
&\quad-{\theta}_{k,i}\bigg) -\(\hat{\theta}_{k-1,ij}-{\theta}_{k-1,i}\)\bigg\|\nonumber\\
\leq&\left\|\gamma b_k\psi_k\left(G_{k,ij}(C_{ij}-\psi_k^T \hat{\theta}_{k-1,ij})-z_{k,ij}\right)\right\|
+\left\|{\theta}_{k,i} -{\theta}_{k-1,i}\right\|\nonumber\\
\leq&(\gamma\bar{\psi}+\beta\bar{\phi}+2\beta N_i\bar{\theta})b_k, \text{for } i\in \mathcal{N}_{i,k}, \nonumber
\end{align}
and hence,
\vspace{-10pt}
\begin{align}
\left\|\varepsilon_{k+l,ij}-\varepsilon_{k,ij}\right\|
\leq&\sum_{p=k+1}^{k+l}\|\varepsilon_{p,ij}-\varepsilon_{p-1,ij}\|\nonumber\\
\leq&(\gamma\bar{\psi}+\beta\bar{\phi}+2\beta N_i\bar{\theta})\sum_{p=k+1}^{k+l}b_p.\nonumber
\end{align}
That is, (\ref{varepsilon}) holds.
Then, from (\ref{UkVk})-(\ref{varepsilon}) and Assumption \ref{Ab}, we have
\vspace{-5pt}
\begin{align}
&\left\|U_{k+l}-U_{k}\right\|=\left\|\mathbb{E}[(\varepsilon_{k+l}-\varepsilon_{k})^T (\varepsilon_{k+l}+\varepsilon_{k})]\right\| =O(b_{k+l}),\nonumber\\
&\left\|V_{k+l}-V_{k}\right\|=\left\|\mathbb{E}[(\tilde{\Theta}_{k+l}-\tilde{\Theta}_{k})^T (\tilde{\Theta}_{k+l}+\tilde{\Theta}_{k})]\right\|=O(b_{k+l}),\nonumber\\
&\left\|\varepsilon_{k+l}-\varepsilon_{k}\right\|=O(b_{k+l}),\nonumber
\end{align}
for all $l\geq 0$.
\qed

\vspace{-5pt}
\subsection{Proof of Lemma \ref{lemmaUk}}\label{appA}
At first, we consider $\varepsilon_{k,ij}$ for the case of $j\in\mathcal{N}_{i,k}$.
From (\ref{AE}), (\ref{AF}) and Lemma \ref{lem:proj},  we have
\vspace{-5pt}
\begin{align}
&\varepsilon_{k,ij}^T\varepsilon_{k,ij}\nonumber\\
\leq &\varepsilon_{k-1,ij}^T\varepsilon_{k-1,ij}
+2\gamma b_k\varepsilon_{k-1,ij}^T\psi_k(\hat{G}_{k,ij}-z_{k,ij})\nonumber\\
&-2\beta b_k\varepsilon_{k-1,ij}^T\phi_{k,j} (\hat{F}_{k,j}-s_{k,j})
+\gamma^2 b_k^2\psi_k^T\psi_k(\hat{G}_{k,ij}-z_{k,ij})^2\nonumber\\
&-2\beta b_k\varepsilon_{k-1,ij}^T\sum_{l\in \mathcal{N}_{j,k}}a_{jl,k} (\hat{\theta}_{k-1,jl}-{\theta}_{k-1,j})\nonumber\\
&-2\gamma\beta b_k^2\psi_k^T\phi_{k,j}(\hat{G}_{k,ij}-z_{k,ij})\cdot (\hat{F}_{k,j}-s_{k,j})\nonumber
\end{align}
\begin{align}\label{appA3}
&-2\gamma\beta b_k^2(\hat{G}_{k,ij}-z_{k,ij})\psi_k^T\sum_{l\in \mathcal{N}_{j,k}}a_{jl,k} (\hat{\theta}_{k-1,jl}-{\theta}_{k-1,j})\nonumber\\
&+2\beta^2 b_k^2 (\hat{F}_{k,j}-s_{k,j})\phi_{k,i}^T\sum_{l\in \mathcal{N}_{j,k}}a_{jl,k} (\hat{\theta}_{k-1,jl}-{\theta}_{k-1,j})\nonumber\\
&+\beta^2 b_k^2 \sum_{l\in \mathcal{N}_{j,k}}a_{jl,k} (\hat{\theta}_{k-1,jl}-{\theta}_{k-1,j})^T
\sum_{l\in \mathcal{N}_{j,k}}a_{jl,k}\nonumber\\
&\cdot (\hat{\theta}_{k-1,jl}-{\theta}_{k-1,j})+\beta^2 b_k^2\phi_{k,j}^T\phi_{k,j}(\hat{F}_{k,j}-s_{k,j})^2\nonumber\\
= &\varepsilon_{k-1,ij}^T\varepsilon_{k-1,ij}+2\gamma b_k\varepsilon_{k-1,ij}^T\psi_k(\hat{G}_{k,ij}-z_{k,ij})\nonumber\\
&-2\beta b_k\varepsilon_{k-1,ij}^T\sum_{l\in \mathcal{N}_{j,k}}a_{jl,k} (\hat{\theta}_{k-1,jl}-{\theta}_{k-1,j})\nonumber\\
&-2\beta b_k\varepsilon_{k-1,ij}^T\phi_{k,j} (\hat{F}_{k,j}-s_{k,j})+O\(b_k^2\), 
\end{align}
where the last equality is got by (\ref{appA1})-(\ref{appA2}) and the boundedness of $\hat{F}_{k,i}$ and $\hat{G}_{k,ij}$, which is got because $F_{k,i}(\cdot)$ and ${G}_{k,ij}(\cdot)$ are continuous functions.

By Assumption \ref{AD} and differential mean value theorem,
\begin{align}\label{appA4}
&\mathbb{E}[2\gamma b_k\varepsilon_{k-1,ij}^T\psi_k(\hat{G}_{k,ij}-z_{k,ij})|\mathcal{F}_{k-1}]\nonumber\\
=&2\gamma b_k\varepsilon_{k-1,ij}^T\psi_k\(\hat{G}_{k,ij}-\mathbb{E}[z_{k,ij}|\mathcal{F}_{k-1}]\)\nonumber\\
=&2\gamma b_k\varepsilon_{k-1,ij}^T\psi_k\big(G_{k,ij}(C_{ij}-\psi^T_k\hat\theta_{k-1,ij}) \nonumber\\ &-G_{k,ij}\(C_{ij}-\psi^T_k\theta_{k-1,j}\)\big)\nonumber\\
=&-2\gamma g_{k,ij}\(\xi_{k,ij}\)b_k\varepsilon_{k-1,ij}^T\psi_k\psi_k^T\varepsilon_{k-1,ij} \nonumber\\
\leq& -2\gamma \underline{g} b_k\varepsilon_{k-1,ij}^T\psi_k\psi_k^T\varepsilon_{k-1,ij}.
\end{align}
where $\xi_{k,ij}$ is in the interval between $C_{ij}-\psi^T_k\hat\theta_{k-1,ij}$ and $C_{ij}-\psi^T_k\theta_{k-1,j}$ such that $G_{k,ij}\(C_{ij}-\psi^T_k\hat\theta_{k-1,ij}\) -G_{k,ij}\(C_{ij}-\psi^T_k\theta_{k-1,j}\)=-g_{k,ij}\(\xi_{k,ij}\)\psi_k^T\varepsilon_{k,ij}$.

By the differential mean value theorem, Cauchy-Schwarz inequality and Assumptions \ref{AR}-\ref{AD},  we have
\begin{align}\label{appA5}
&-\mathbb{E}[2\beta b_k\varepsilon_{k-1,ij}^T\phi_{k,j} (\hat{F}_{k,j}-s_{k,j})|\mathcal{F}_{k-1}]\nonumber\\
=&-2\beta b_k\varepsilon_{k-1,ij}^T\phi_{k,j} \(\hat{F}_{k,j}-\mathbb{E}[s_{k,j}|\mathcal{F}_{k-1}]\)\nonumber\\
=&-2\beta b_k\varepsilon_{k-1,ij}^T\phi_{k,j} \(F_{k,j}(C_j-\phi^T_{k,j}\theta_{k-1,j}) -F_{k,j}(C_j-\phi^T_{k,j}\theta)\)\nonumber\\
=&2\beta b_k\varepsilon_{k-1,ij}^T\phi_{k,j}\phi^T_{k,j}\tilde\theta_{k-1,j}f_{k,j}\(\zeta_{k,j}\)\nonumber\\
\leq& 2\beta \bar{f} \bar{\phi}^2 b_k \|\varepsilon_{k-1,ij}\|\|\tilde\theta_{k-1,j}\|\nonumber\\
\leq& \beta \bar{f} \bar{\phi}^2 b_k \(\frac{1}{\alpha_1}\|\varepsilon_{k-1,ij}\|^2 +\alpha_1\|\tilde\theta_{k-1,j}\|^2\),
\end{align}
where $\zeta_{k,j}$ is in the interval between $C_j-\phi^T_{k,j}\theta_{k-1,j}$ and $C_j-\phi^T_{k,j}\theta$ such that $F_{k,j}(C_j-\phi^T_{k,j}\theta_{k-1,j}) -F_{k,j}(C_j-\phi^T_{k,j}\theta) =-f_{k,j}\(\zeta_{k,j}\)\phi^T_{k,j}\tilde\theta_{k-1,j}$, and $\alpha_1\in(0,\infty)$.

Substituting (\ref{appA4}) and (\ref{appA5}) into (\ref{appA3}) and  calculating its mathematical expectation, we can obtain for $j\in\mathcal{N}_{i,k}$
\vspace{-5pt}
\begin{align}\label{appA6}
&\mathbb{E}\varepsilon_{k,ij}^T\varepsilon_{k,ij}\nonumber\\
\leq &\mathbb{E}\varepsilon_{k-1,ij}^T\varepsilon_{k-1,ij}+2\gamma b_k\mathbb{E}\varepsilon_{k-1,ij}^T\psi_k(\hat{G}_{k,ij}-z_{k,ij})\nonumber\\
&-2\beta b_k\mathbb{E}\varepsilon_{k-1,ij}^T\sum_{l\in \mathcal{N}_{j,k}}a_{jl,k} (\hat{\theta}_{k-1,jl}-{\theta}_{k-1,j})\nonumber\\
&-2\beta b_k\mathbb{E}\varepsilon_{k-1,ij}^T\phi_{k,j} (\hat{F}_{k,j}-s_{k,j})+O\(b_k^2\)\nonumber\\
\leq  &\mathbb{E}\varepsilon_{k-1,ij}^T\varepsilon_{k-1,ij}-2\gamma \underline{g} b_k\mathbb{E}\varepsilon_{k-1,ij}^T\psi_k\psi_k^T\varepsilon_{k-1,ij}\nonumber\\
&+ \frac{\beta \bar{f} \bar{\phi}^2}{\alpha_1}b_k \mathbb{E}\|\varepsilon_{k-1,ij}\|^2
+\alpha_1\beta \bar{f} \bar{\phi}^2\mathbb{E}\|\tilde\theta_{k-1,j}\|^2 \nonumber\\
&-2\beta b_k\mathbb{E}\varepsilon_{k-1,ij}^T\sum_{l\in \mathcal{N}_{j,k}}a_{jl,k} ({\theta}_{k-1,l}-{\theta}_{k-1,j})\nonumber\\
&-2\beta b_k\mathbb{E}\varepsilon_{k-1,ij}^T\sum_{l\in \mathcal{N}_{j,k}}a_{jl,k}\varepsilon_{k,jl}+O\(b_k^2\).
\end{align}
Next, we consider $\varepsilon_{k,ij}$ for the case of $j\in\mathcal{N}_{i}^{\mathcal{C}}\setminus \mathcal{N}_{i,k}$. By (\ref{AE}), (\ref{AF}) and Lemma \ref{lem:proj}, we can obtain
\vspace{-5pt}
\begin{align}\label{appA7}
&\varepsilon_{k,ij}^T\varepsilon_{k,ij}\nonumber\\
\leq&\big(\varepsilon_{k-1,ij}-\beta b_k\phi_{k,j} (\hat{F}_{k,j}-s_{k,j})\nonumber\\
&\quad-\beta b_k\sum_{l\in \mathcal{N}_{j,k}}a_{jl,k} (\hat{\theta}_{k-1,jl}-{\theta}_{k-1,j})\big)^T\nonumber\\
&\cdot \big(\varepsilon_{k-1,ij}-\beta b_k\phi_{k,j} (\hat{F}_{k,j}-s_{k,j})\nonumber\\
&\quad-\beta b_k\sum_{l\in \mathcal{N}_{j,k}}a_{jl,k} (\hat{\theta}_{k-1,jl}-{\theta}_{k-1,j})\big)\nonumber\\
= &\varepsilon_{k-1,ij}^T\varepsilon_{k-1,ij}
-2\beta b_k\varepsilon_{k-1,ij}^T\phi_{k,j} (\hat{F}_{k,j}-s_{k,j})+O\(b_k^2\)\nonumber\\
&-2\beta b_k\varepsilon_{k-1,ij}^T\sum_{l\in \mathcal{N}_{j,k}}a_{jl,k} (\hat{\theta}_{k-1,jl}-{\theta}_{k-1,j}),
\end{align}
where the last equality is got similarly to (\ref{appA3}) based on (\ref{appA1})-(\ref{appA2}) and the boundedness of $\hat{F}_{k,i}$.

Taking (\ref{appA5}) into (\ref{appA7}) and calculating its mathematical expectation, we can get for $j\in \mathcal{N}_{i}^{\mathcal{C}}\setminus \mathcal{N}_{i,k}$
\vspace{-5pt}
\begin{align}\label{appA8}
&\mathbb{E}\varepsilon_{k,ij}^T\varepsilon_{k,ij}\nonumber\\
\leq  &\mathbb{E}\varepsilon_{k-1,ij}^T\varepsilon_{k-1,ij}
+ \frac{\beta \bar{f} \bar{\phi}^2}{\alpha_1}b_k \mathbb{E}\|\varepsilon_{k-1,ij}\|^2 +O\(b_k^2\)\nonumber\\
&+\alpha_1\beta \bar{f} \bar{\phi}^2\mathbb{E}\|\tilde\theta_{k-1,j}\|^2
-2\beta b_k\mathbb{E}\varepsilon_{k-1,ij}^T\sum_{l\in \mathcal{N}_{j,k}}a_{jl,k}\varepsilon_{k,jl} \nonumber\\
&-2\beta b_k\mathbb{E}\varepsilon_{k-1,ij}^T\sum_{l\in \mathcal{N}_{j,k}}a_{jl,k} ({\theta}_{k-1,l}-{\theta}_{k-1,j}).
\end{align}
\vspace{-5pt}
By (\ref{EFQDIA}), (\ref{UkVk}), (\ref{appA6}), (\ref{appA8}) and the notations in Section \ref{sec:nota},
\vspace{-5pt}
\begin{align}
U_k\leq & U_{k-1}-2\gamma \underline{g} b_k\mathbb{E}\varepsilon_{k-1}^T\Psi_kR_k\Psi_k^T\varepsilon_{k-1}
+ \frac{\beta \bar{f} \bar{\phi}^2}{\alpha_1}b_k \mathbb{E}\|\varepsilon_{k-1}\|^2\nonumber\\
&+\alpha_1\beta \bar{f} \bar{\phi}^2b_k\mathbb{E}\tilde\Theta_{k-1}^TQ^T\otimes I_n Q\otimes I_n\tilde\Theta_{k-1}\nonumber\\
&-2\beta b_k\mathbb{E}\varepsilon_{k-1}^TQ\otimes I_n P_k\otimes I_n\varepsilon_{k-1}\nonumber\\
&-2\beta b_k\mathbb{E}\varepsilon_{k-1}^TQ\otimes I_n \mathcal{L}_k\otimes I_n\tilde\Theta_{k-1}+O\(b_k^2\)\nonumber\\
\leq & \(1+{\beta \bar{f} \bar{\phi}^2}b_k/{\alpha_1}\)U_{k-1}-2\gamma \underline{g} b_k\mathbb{E}\varepsilon_{k-1}^T\Psi_kR_k\Psi_k^T\varepsilon_{k-1}\nonumber
\end{align}
\begin{align}\label{appA9}
&-2\beta b_k\mathbb{E}\varepsilon_{k-1}^TQ P_k\otimes I_n\varepsilon_{k-1}
+\frac{\beta}{\alpha_2} b_k\mathbb{E}\varepsilon_{k-1}^TQQ^T\otimes I_n \varepsilon_{k-1}\nonumber\\
&+\alpha_1\beta \bar{f} \bar{\phi}^2b_k\mathbb{E}\tilde\Theta_{k-1}^TQ^TQ\otimes I_n\tilde\Theta_{k-1}\nonumber\\
&+\beta\alpha_2 b_k\mathbb{E}\tilde\Theta_{k-1}\mathcal{L}_k^T\mathcal{L}_k\otimes I_n\tilde\Theta_{k-1}+O\(b_k^2\).
\end{align}
Based on Lemma \ref{LemmaRPQ}, the second term in the right side of (\ref{appA9}) can be estimated as
\vspace{-10pt}
\begin{align}\label{appA92}
&-2\gamma \underline{g} b_k\mathbb{E}\varepsilon_{k-1}^T\Psi_kR_k\Psi_k^T\varepsilon_{k-1} \nonumber\\
=&-2\gamma \underline{g} b_k \mathbb{E}[\mathbb{E}[\varepsilon_{k-1}^T\Psi_kR_k\Psi_k^T\varepsilon_{k-1}|\mathcal{F}_{k-1}]]\nonumber\\
=&-2\gamma \underline{g} b_k \mathbb{E}[\varepsilon_{k-1}^T\Psi_k\mathbb{E}[R_k]\Psi_k^T\varepsilon_{k-1}]\nonumber\\
\leq& -2\gamma \underline{g}\underline{\pi} b_k \mathbb{E}\varepsilon_{k-1}^T\Psi_k\Psi_k^T\varepsilon_{k-1}+O(\lambda_{q}^k),
\end{align}
where the last inequality is got by the boundedness of $\Psi_k$ and $\varepsilon_{k-1}$.
From Assumption \ref{AG}, we learn $\mathcal{L}_k\in\{\mathcal{L}^{(1)},\ldots,\mathcal{L}^{(s)}\}$ and
\vspace{-5pt}
\begin{align}\label{Lk}
\|\mathcal{L}_k\|\leq \max_{1\leq u\leq s}\|\mathcal{L}^{(u)}\|\triangleq \lambda_m.
\end{align}
\vspace{-10pt}
From Lemmas \ref{LemmaRPQ}, taking (\ref{appA92}) and (\ref{Lk}) into (\ref{appA9}) yields
\vspace{-3pt}
\begin{align}\label{appA10}
U_k\leq & \(1+\beta\({\bar{f} \bar{\phi}^2}/{\alpha_1}+2\bar{N}+{\bar{N}}/{\alpha_2}\)b_k\)U_{k-1}\nonumber\\
&-2\gamma \underline{g}\underline{\pi} b_k \mathbb{E}\varepsilon_{k-1}^T\Psi_k\Psi_k^T\varepsilon_{k-1}\nonumber\\
&+\beta(\alpha_1\bar{f}\bar{N} \bar{\phi}^2+\alpha_2\lambda_m^2) b_kV_{k-1}+O\(b_k^2\)\nonumber\\
\leq & U_{k-h}+\beta\(\frac{\bar{f} \bar{\phi}^2}{\alpha_1}+2\bar{N}+\frac{\bar{N}}{\alpha_2}\)\sum_{l=k-h}^{k-1}b_{l+1}U_{l}\nonumber\\
&-2\gamma \underline{g}\underline{\pi} \sum_{l=k-h}^{k-1} \mathbb{E}\varepsilon_{l}^T b_{l+1}\Psi_{l+1}\Psi_{l+1}^T\varepsilon_{l}+O\(b_k^2\)\nonumber\\
&+\beta(\alpha_1\bar{f}\bar{N} \bar{\phi}^2+\alpha_2\lambda_m^2) \sum_{l=k-h}^{k-1}b_{l+1}V_{l}.
\end{align}
By Remark \ref{RAb}, there exists $c_{h}$ such that $c_{h}\triangleq\sup_{|p-q|<h}\frac{b_{p}}{b_{q}}<\infty$. So we have $b_{l+1}\geq \frac{1}{hc_h}\sum_{l=k-h}^{k-1} b_{l+1}$. Then, based on the encoding rule in Algorithm \ref{Algorithm} and Assumption \ref{Ab}, we have
\vspace{-7pt}
\begin{align}\label{appA11}
\sum_{l=k-h}^{k-1} b_{l+1} \Psi_{l+1}\Psi_{l+1}^T
=&\frac{1}{hc_h}\sum_{l=k-h}^{k-1} b_{l+1}\sum_{l=k-h}^{k-1} \Psi_{l+1}\Psi_{l+1}^T\nonumber\\
\geq& \frac{\delta_{\psi}^2}{c_h}\sum_{l=k-h}^{k-1} b_{l+1}I_{nN}.
\end{align}
\vspace{-7pt}
From Lemma \ref{LemmaUVE} and (\ref{appA10})-(\ref{appA11}), we have
\vspace{-24pt}
\begin{small}
\begin{align}
U_k\leq & \(1+\beta\(\frac{\bar{f} \bar{\phi}^2}{\alpha_1}+2\bar{N}+\frac{\bar{N}}{\alpha_2}\)\sum_{l=k-h}^{k-1}b_{l+1}\)U_{k-h}\nonumber\\
&-2\gamma \underline{g}\underline{\pi} \mathbb{E}\varepsilon_{k-h}^T\big(\sum_{l=k-h}^{k-1} b_{l+1} \Psi_{l+1}\Psi_{l+1}^T\big)\varepsilon_{k-h}\nonumber\\
&+\beta(\alpha_1\bar{f}\bar{N} \bar{\phi}^2+\alpha_2\lambda_m^2) \sum_{l=k-h}^{k-1}b_{l+1}V_{k-h} +O\(b_k^2\)\nonumber
\end{align}
\end{small}
\vspace{-30pt}
\begin{small}
\begin{align}
\leq & \(1-\(\frac{2\gamma \underline{g}\underline{\pi}\delta_{\psi}^2}{c_h}-\beta\(\frac{\bar{f} \bar{\phi}^2}{\alpha_1}+2\bar{N}+\frac{\bar{N}}{\alpha_2}\)\)\sum_{l=k-h}^{k-1}b_{l+1}\)U_{k-h}\nonumber\\
&+\beta(\alpha_1\bar{f}\bar{N} \bar{\phi}^2+\alpha_2\lambda_m^2) \sum_{l=k-h}^{k-1}b_{l+1}V_{k-h} +O\(b_k^2\). \nonumber
\end{align}
\end{small}
This completes the proof.\qed

\vspace{-5pt}
\subsection{Proof of Lemma \ref{lemmaV_k}}
From (\ref{AE}), (\ref{appA1})-(\ref{appA2}), Lemma \ref{lem:proj} and the boundedness of $\hat{F}_{k,i}$, we have
\vspace{-5pt}
\begin{align}\label{lemmaV_k1}
&\tilde{\theta}_{k,i}^T\tilde{\theta}_{k,i}\nonumber\\
\leq&\tilde{\theta}_{k-1,i}^T\tilde{\theta}_{k-1,i}+2\beta b_k \tilde{\theta}_{k-1,i}^T\phi_{k,i} (\hat{F}_{k,i}-s_{k,i})\nonumber\\
&+2\beta b_k \tilde{\theta}_{k-1,i}^T\sum_{j\in \mathcal{N}_{i,k}}a_{ij,k} (\hat{\theta}_{k-1,ij}-{\theta}_{k-1,i})\nonumber\\
&+2\beta^2 b_k^2(\hat{F}_{k,i}-s_{k,i})\phi_{k,i}^T\sum_{j\in \mathcal{N}_{i,k}}a_{ij,k} (\hat{\theta}_{k-1,ij}-{\theta}_{k-1,i})\nonumber\\
&+\beta^2 b_k^2\bigg(\sum_{j\in \mathcal{N}_{i,k}}a_{ij,k} (\hat{\theta}_{k-1,ij}-{\theta}_{k-1,i})\bigg)^T
\sum_{j\in \mathcal{N}_{i,k}}a_{ij,k} \nonumber\\
&\cdot(\hat{\theta}_{k-1,ij}-{\theta}_{k-1,i})+\beta^2 b_k^2\phi_{k,i}^T\phi_{k,i} (\hat{F}_{k,i}-s_{k,i})^2\nonumber\\
\leq&\tilde{\theta}_{k-1,i}^T\tilde{\theta}_{k-1,i}
+2\beta b_k \tilde{\theta}_{k-1,i}^T\phi_{k,i} (\hat{F}_{k,i}-s_{k,i})+O(b_k^2)\nonumber\\
&+2\beta b_k \tilde{\theta}_{k-1,i}^T\sum_{j\in \mathcal{N}_{i,k}}a_{ij,k} (\hat{\theta}_{k-1,ij}-{\theta}_{k-1,i}).
\end{align}
\vspace{-5pt}
By Assumption \ref{AD} and differential mean value theorem,
\vspace{-5pt}
\begin{align}\label{lemmaV_k2}
&\mathbb{E}[2\beta b_k\tilde{\theta}_{k-1,i}^T\phi_{k,i} (\hat{F}_{k,i}-s_{k,i})|\mathcal{F}_{k-1}]\nonumber\\
=&2\beta b_k\tilde{\theta}_{k-1,i}^T\phi_{k,i} \(\hat{F}_{k,i}-\mathbb{E}[s_{k,i}|\mathcal{F}_{k-1}]\)\nonumber\\
=&2\beta b_k\tilde{\theta}_{k-1,i}^T\phi_{k,i} \(F_{k,i}(C_i-\phi^T_{k,i}\theta_{k-1,i}) -F_{k,i}(C_i-\phi^T_{k,i}\theta)\)\nonumber\\
=&-2\beta f_{k,i}\(\zeta_{k,i}\)b_k\tilde{\theta}_{k-1,i}^T\phi_{k,i}\phi^T_{k,i}\tilde\theta_{k-1,i}\nonumber\\
\leq& -2\beta \underline{f} b_k\tilde{\theta}_{k-1,i}^T\phi_{k,i}\phi^T_{k,i}\tilde\theta_{k-1,i},
\end{align}
where $\zeta_{k,i}$ is defined as (\ref{appA5}).

Substituting (\ref{lemmaV_k2}) into (\ref{lemmaV_k1}) and  calculating its mathematical expectation, we can obtain
\vspace{-5pt}
\begin{align}\label{lemmaV_k3}
&\mathbb{E}\tilde{\theta}_{k,i}^T\tilde{\theta}_{k,i}\nonumber\\
\leq&\mathbb{E}\tilde{\theta}_{k-1,i}^T\tilde{\theta}_{k-1,i}
-2\beta b_k \underline{f}\mathbb{E}\tilde{\theta}_{k-1,i}^T\phi_{k,i}\phi^T_{k,i}\tilde\theta_{k-1,i}\nonumber\\
&+2\beta b_k \mathbb{E}\tilde{\theta}_{k-1,i}^T\sum_{j\in \mathcal{N}_{i,k}}a_{ij,k} (\hat{\theta}_{k-1,ij}-{\theta}_{k-1,i})+O(b_k^2)\nonumber\\
\leq&\mathbb{E}\tilde{\theta}_{k-1,i}^T\tilde{\theta}_{k-1,i}
-2\beta b_k \underline{f}\mathbb{E}\tilde{\theta}_{k-1,i}^T\phi_{k,i}\phi^T_{k,i}\tilde\theta_{k-1,i}\nonumber\\
&+2\beta b_k \mathbb{E}\tilde{\theta}_{k-1,i}^T\sum_{j\in \mathcal{N}_{i,k}}a_{ij,k} (\tilde{\theta}_{k-1,j}-\tilde{\theta}_{k-1,i})\nonumber\\
&+2\beta b_k \mathbb{E}\tilde{\theta}_{k-1,i}^T\sum_{j\in \mathcal{N}_{i,k}}a_{ij,k}\varepsilon_{k-1,ij}+O(b_k^2).
\end{align}
Based on (\ref{UkVk}), (\ref{lemmaV_k3}), Lemma \ref{LemmaRPQ} and the notations in Section \ref{sec:nota}, we have
\vspace{-7pt}
\begin{align}\label{lemmaV_k4}
V_k=&\mathbb{E}\tilde{\Theta}_{k}^T\tilde{\Theta}_{k}
\leq V_{k-1}-2\beta b_k \underline{f}\mathbb{E}\tilde{\Theta}_{k-1}^T\Phi_{k}\Phi^T_{k}\tilde\Theta_{k-1}\nonumber\\
&+2\beta b_k \mathbb{E}\tilde{\Theta}_{k-1}^TP_k\otimes I_n\varepsilon_{k-1}\nonumber\\
&-2\beta b_k \mathbb{E}\tilde{\Theta}_{k-1}^T\mathcal{L}_k\otimes I_n\tilde{\Theta}_{k-1}+O(b_k^2)\nonumber\\
\leq& V_{k-1}-2\beta b_k \underline{f}\mathbb{E}\tilde{\Theta}_{k-1}^T\Phi_{k}\Phi^T_{k}\tilde\Theta_{k-1}\nonumber\\
&+2\beta b_k \mathbb{E}\tilde{\Theta}_{k-1}^T\mathcal{L}_k\otimes I_n\tilde{\Theta}_{k-1}+O(b_k^2)\nonumber\\
&+{\beta }/{\alpha_3}b_k \mathbb{E}\tilde{\Theta}_{k-1}^T\tilde{\Theta}_{k-1}
+{\alpha_3}\beta\bar{N} b_k U_{k-1} \nonumber\\
\leq&V_{k-h}+\frac{\beta }{\alpha_3}\sum_{l=k-h}^{k-1}b_{l+1}V_{l} +{\alpha_3}\beta\bar{N}\sum_{l=k-h}^{k-1}b_{l+1}U_{l}\nonumber\\
&-2\beta \sum_{l=k-h}^{k-1}b_{l+1}\mathbb{E}\tilde{\Theta}_{l}^T\mathcal{L}_{l+1}\otimes I_n\tilde{\Theta}_{l}\nonumber\\
&-2\beta\underline{f}\sum_{l=k-h}^{k-1}b_{l+1}\mathbb{E}\tilde{\Theta}_{l}^T\Phi_{l+1}\Phi^T_{l+1}\tilde\Theta_{l}
+O(b_k^2).
\end{align}
\vspace{-7pt}
By Remark \ref{RLk}, we have
\vspace{-7pt}
\begin{align}\label{ELk}
\mathbb{E}\mathcal{L}_{k}=\sum_{u=1}^s\pi\mathcal{L}^{(u)}+O(\lambda_{a}^k)={\mathcal{L}}_{\mathcal{C}}+O(\lambda_{a}^k).
\end{align}
\vspace{-7pt}
From Assumption \ref{Ab}, Lemma \ref{LemmaUVE} and (\ref{lemmaV_k4}),  we have
\vspace{-7pt}
\begin{align}\label{lemmaV_k5}
V_k
\leq&V_{k-h}+\frac{\beta }{\alpha_3}\sum_{l=k-h}^{k-1}b_{l+1}V_{k-h}+\frac{\beta }{\alpha_3} \sum_{l=k-h}^{k-1}b_{l+1}(V_l-V_{k-h}) \nonumber\\
&-2\beta \underline{f}\mathbb{E}\tilde{\Theta}_{k-h}^T\sum_{l=k-h}^{k-1}b_{l+1}\Phi_{l+1}\Phi^T_{l+1} \tilde\Theta_{k-h}\nonumber\\
&-2\beta \underline{f} \sum_{l=k-h}^{k-1}b_{l+1}\mathbb{E}\tilde{\Theta}_{l}^T \Phi_{l+1}\Phi^T_{l+1}(\tilde\Theta_{l}-\tilde\Theta_{k-h})\nonumber\\
&-2\beta \mathbb{E}\tilde{\Theta}_{k-h}^T\sum_{l=k-h}^{k-1}b_{l+1}\mathcal{L}_{l+1}\otimes I_n\tilde{\Theta}_{k-h}+O(b_k^2)\nonumber\\
&-2\beta \sum_{l=k-h}^{k-1}b_{l+1}\mathbb{E}\tilde{\Theta}_{l}^T\mathcal{L}_{l+1}\otimes I_n(\tilde{\Theta}_{l}-\tilde{\Theta}_{k-h})\nonumber\\
&+{\alpha_3}\beta\bar{N}\sum_{l=k-h}^{k-1}b_{l+1}U_{k-h} +{\alpha_3}\beta\bar{N}\sum_{l=k-h}^{k-1}b_{l+1}(U_{l}-U_{k-h}) \nonumber\\
\leq&V_{k-h}+\frac{\beta }{\alpha_3}\sum_{l=k-h}^{k-1}b_{l+1}V_{k-h}\nonumber\\
&-\frac{2\beta \underline{f}}{hc_h}\sum_{l=k-h}^{k-1}b_{l+1}\mathbb{E}\tilde{\Theta}_{k-h}^T \sum_{l=k-h}^{k-1}\Phi_{l+1}\Phi^T_{l+1} \tilde\Theta_{k-h}\nonumber\\
&-2\beta \mathbb{E}\tilde{\Theta}_{k-h}^T\sum_{l=k-h}^{k-1}b_{l+1}\mathcal{L}_{l+1}\otimes I_n\tilde{\Theta}_{k-h}\nonumber\\
&+{\alpha_3}\beta\bar{N}\sum_{l=k-h}^{k-1}b_{l+1}U_{k-h}+O(b_k^2).
\end{align}
Then from Assumption \ref{Ab}, Lemma \ref{LemmaUVE}, and (\ref{lemmaV_k4})-(\ref{ELk}),
\vspace{-10pt}
\begin{align}\label{ELk2}
&2\beta \mathbb{E}\tilde{\Theta}_{k-h}^T\sum_{l=k-h}^{k-1}b_{l+1}\mathcal{L}_{l+1}\otimes I_n\tilde{\Theta}_{k-h}\nonumber\\
=&2\beta \mathbb{E}\left[\tilde{\Theta}_{k-h}^T\sum_{l=k-h}^{k-1}b_{l+1}\mathbb{E}[\mathcal{L}_{l+1}\otimes I_n|\mathcal{F}_{k-h}]\tilde{\Theta}_{k-h}\right]\nonumber\\
=&2\beta \mathbb{E}\left[\tilde{\Theta}_{k-h}^T\sum_{l=k-h}^{k-1}b_{l+1}\mathbb{E}[\mathcal{L}_{l+1}\otimes I_n]\tilde{\Theta}_{k-h}\right]\nonumber\\
=&2\beta \sum_{l=k-h}^{k-1}b_{l+1}\mathbb{E}\tilde{\Theta}_{k-h}^T{\mathcal{L}}_{\mathcal{C}}\otimes I_n\tilde{\Theta}_{k-h}
+O(\lambda^k_ab_k)\nonumber\\
=&\beta \sum_{l=k-h}^{k-1}b_{l+1}\mathbb{E}\tilde{\Theta}_{k-h}^T({\mathcal{L}}_{\mathcal{C}}+{\mathcal{L}}_{\mathcal{C}}^T)\otimes I_n\tilde{\Theta}_{k-h}+O(\lambda^k_ab_k)\nonumber\\
=&2\beta \sum_{l=k-h}^{k-1}b_{l+1}\mathbb{E}\tilde{\Theta}_{k-h}^T\hat{\mathcal{L}}_{\mathcal{C}}\otimes I_n\tilde{\Theta}_{k-h}
+O(\lambda^k_ab_k),
\end{align}
where $\hat{\mathcal{L}}_{\mathcal{C}}$ is the Laplacian matrix of the mirror graph of the union graph $\mathcal{G}_{\mathcal{C}}$. By Lemma \ref{lem:hatL} and Assumption \ref{AG}, the smallest nonzero eigenvalue $\hat{\lambda}_2$ of $\hat{\mathcal{L}}_{\mathcal{C}}$ satisfies $\hat{\lambda}_2>0$.
From Assumption \ref{AR}, Lemma \ref{lemmaAM}, and (\ref{ELk2}), we have
\vspace{-10pt}
\begin{align}\label{lemmaV_k6}
&-\frac{\underline{f}}{h}\sum_{l=k-h}^{k-1}b_{l+1}\mathbb{E}\tilde{\Theta}_{k-h}^T \sum_{l=k-h}^{k-1}\Phi_{l+1}\Phi^T_{l+1} \tilde\Theta_{k-h}\nonumber\\
&-\mathbb{E}\tilde{\Theta}_{k-h}^T\sum_{l=k-h}^{k-1}b_{l+1}\mathcal{L}_{l+1}\otimes I_n\tilde{\Theta}_{k-h}\nonumber\\
\leq&-\sum_{l=k-h}^{k-1}b_{l+1}\mathbb{E}\tilde{\Theta}_{k-h}^T \bigg(\frac{\underline{f}}{h}  \mathbb{E}\left[\sum_{l=k-h}^{k-1}\Phi_{l+1}\Phi^T_{l+1}\bigg|\mathcal{F}_{k-h}\right]\nonumber\\
&+ \hat{\mathcal{L}}_{\mathcal{C}}\otimes I_n\bigg)\tilde\Theta_{k-h}+O(\lambda^k_ab_k)\nonumber\\
\leq&-\frac{\sigma}{m\underline{f}\delta_{\phi}^2}\mathbb{E}\lambda_{\min}\( \mathbb{E} \left[\frac{\underline{f}}{h}\sum_{i=1}^m\sum_{l=k-h}^{k-1}\phi_{l+1,i}\phi^T_{l+1,i}\bigg| \mathcal{F}_{k-h}\right]\) \nonumber\\
&\cdot\sum_{l=k-h}^{k-1}b_{l+1}\tilde{\Theta}_{k-h}^T \tilde\Theta_{k-h}+O(\lambda^k_ab_k)\nonumber\\
\leq&-\sigma\sum_{l=k-h}^{k-1}b_{l+1}V_{k-h}+O(\lambda^k_ab_k),
\end{align}
\vspace{-10pt}
where $\sigma=\frac{h\underline{f}\hat{\lambda}_{2}\delta_{\phi}^2}{2\underline{f}\bar{\phi}^2+h\hat{\lambda}_{2}}$.
Taking (\ref{lemmaV_k6}) into (\ref{lemmaV_k5}) gives
\begin{align}
V_k
\leq&\(1-\(\frac{2\beta\sigma}{c_h}-\frac{\beta }{\alpha_3}\)\sum_{l=k-h}^{k-1}b_{l+1}\)V_{k-h}\nonumber\\
&+{\alpha_3}\beta\bar{N}\sum_{l=k-h}^{k-1}b_{l+1}U_{k-h}+O(b_k^2).\nonumber
\end{align}
\vspace{-10pt}
This completes the proof.\qed

\vspace{-5pt}
\subsection{Proof of Theorem \ref{thm_c}}
Based on Lemmas \ref{lemmaUk} and \ref{lemmaV_k}, taking $\alpha_1={c_h}/(\sigma\bar{f}\bar{\phi}^2)$,  $\alpha_2={c_h\bar{N}}/{\sigma\lambda_{m}^2}$ and $\alpha_3={2c_h}/{\sigma}$ can give
\vspace{-5pt}
\begin{align}\label{thm_c1}
\left\{
   \begin{array}{ll}
U_k\leq\bigg(1-\(\frac{2\gamma \underline{g}\underline{\pi}\delta_{\psi}^2}{c_h}-\beta\(\frac{\bar{f}^2 \bar{\phi}^4\sigma}{c_h}+2\bar{N}+\frac{\sigma\lambda_{m}^2}{c_h}\)\)\sum_{l=k-h}^{k-1}\\
\qquad \cdot b_{l+1}\bigg)U_{k-h}
+\frac{2\beta c_h\bar{N}}{\sigma}\sum_{l=k-h}^{k-1}b_{l+1}V_{k-h} +O\(b_k^2\),\\
V_k\leq\(1-\frac{3\beta\sigma}{2c_h}\sum_{l=k-h}^{k-1}b_{l+1}\)V_{k-h}\\
  \qquad+\frac{2\beta c_h \bar{N}}{\sigma}\sum_{l=k-h}^{k-1}b_{l+1}U_{k-h}+O\(b_{k}^2\),
   \end{array}
\right.
\end{align}
Take $w_1=\frac{2\gamma \underline{g}\underline{\pi}\delta_{\psi}^2}{c_h}-\beta\(\frac{\bar{f}^2 \bar{\phi}^4\sigma}{c_h}+2\bar{N}+\frac{\sigma\lambda_{m}^2}{c_h}\)$, $w_2=-\frac{2\beta c_h\bar{N}}{\sigma}$ and $w_3=\frac{3\beta\sigma}{2c_h}$.
Denote
\vspace{-10pt}
$$
Z_k=\begin{bmatrix}
U_{k}\\
V_{k}
\end{bmatrix}, \quad
W=\begin{bmatrix}
w_{1}&w_{2}\\
w_{2}&w_{3}
\end{bmatrix}.
$$
By Assumption \ref{Ab}, there exists $K>0$ such that $\sum_{l=k-h}^{k-1}b_{l+1}\lambda_{\max}(W)>1$ for $k>K$.
From (\ref{thm_c1}),
\vspace{-10pt}
\begin{align}\label{thm_c2}
&\|Z_k\|\leq\big\|I_2-W\sum_{l=k-h}^{k-1}b_{l+1}\big\|\|Z_{k-h}\| +O\(b_k^2\)\nonumber\\
\leq&\big(1-\lambda_{\min}(W)\sum_{l=k-h}^{k-1}b_{l+1}\big)\|Z_{k-h}\| +O\(b_k^2\),
\end{align}
for all $k>K$.
Actually, in order to prove the convergence of $U_k$ and $V_k$, we only need to prove
$\lim_{k\rightarrow\infty}\|Z_k\|=0$.
Then, based on Lemma \ref{lem:xl} and Assumption \ref{Ab}, we only need to prove that $\lambda_{\min}(W)>0$
because of $\sum_{k=1}^{\infty}b_{k}=\infty$ and $\lim_{k \rightarrow \infty}{b_k^2}/{(\sum_{l=k-h}^{k-1}b_{l+1})}=0$.

Let $|\lambda I_2-W|=(\lambda-w_{1})(\lambda-w_{3})-w_{2}^{2}=0$. Then,
$$\lambda_{\min}(W)=\frac{1}{2}\Big(w_{1}+w_{3}-\sqrt{(w_{1}+w_{3})^{2}-4(w_{1}w_{3}-w_{2}^{2})} \Big).$$
If  $\gamma>\frac{\beta}{2\underline{g}\underline{\pi}\delta_{\psi}^2}
\big(2c_h\bar{N}+\sigma\(\bar{f}^2 \bar{\phi}^4+\lambda_{m}^2\)+\frac{8c_h^4\bar{N}^2}{3\sigma^3} \big)$,
then we can obtain $w_{1}w_{3}>w_{2}^{2}$ and $w_{1}>0$, which implies $\lambda_{\min}(W)>0$.
Then, from (\ref{UkVk}), we obtain $\lim_{k\rightarrow\infty}\mathbb{E}\|\tilde{\Theta}_k\|^2=0$ and $
\lim_{k\rightarrow\infty}\mathbb{E}\|\varepsilon_k\|^2=0.$

\vspace{-5pt}
\subsection{Proof of Theorem \ref{thm_cr}}
\vspace{-5pt}
\textbf{Part I}: The case of $b_k=\frac{1}{k^p},p\in(1/2,1)$.\\
Based on (\ref{thm_c2}), we have
\vspace{-5pt}
\begin{align}
\|Z_k\|
\leq&\left(1-\lambda_{\min}(W)\sum_{l=k-h}^{k-1}\frac{1}{(l+1)^p}\right)\|Z_{k-h}\| +O\(\frac{1}{k^{2p}}\),\nonumber\\
\leq&\left(1-\frac{\lambda_{\min}(W)h}{k^p}\right)\|Z_{k-h}\| +O\(\frac{1}{k^{2p}}\),\nonumber
\end{align}
\begin{align}
\leq&\prod^{\left\lfloor\frac{k-K}{h}\right\rfloor-1}_{l=0}\left(1-\frac{\lambda_{\min}(W)h}{(k-lh)^p}\right)
\left\|Z_{k-\left\lfloor\frac{k-K}{h}\right\rfloor h}\right\|\nonumber\\
&+\sum^{\left\lfloor\frac{k-K}{h}\right\rfloor}_{l=1}
\prod^{l-1}_{q=0}\left(1-\frac{\lambda_{\min}(W)h}{(k-qh)^p}\right)O\left(\frac{1}{(k-lh)^{2p}}\right)\nonumber\\
\leq&\prod^{\left\lceil\frac{k}{h}\right\rceil}_{l=\left\lceil\frac{K}{h}\right\rceil+\kappa+1}
\left(1-\frac{\lambda_{\min}(W)h}{(lh)^{p}}\right)
\left\|Z_{k-\left\lfloor\frac{k-K}{h}\right\rfloor h}\right\|\nonumber\\
&+\sum^{\left\lfloor\frac{k}{h}\right\rfloor-1}_{l=\left\lceil\frac{K}{h}\right\rceil+1}
\prod^{\left\lceil\frac{k}{h}\right\rceil}_{q=\left\lceil\frac{K}{h}\right\rceil+\kappa+l+1}
\left(1-\frac{\lambda_{\min}(W)h}{(qh)^{p}}\right)O\left(\frac{1}{(lh)^{2p}}\right)\nonumber\\
\leq&\prod^{\left\lceil\frac{k}{h}\right\rceil}_{l=\left\lceil\frac{K}{h}\right\rceil+\kappa+1}
\left(1-\frac{\lambda_{\min}(W)h^{1-p}}{l^{p}}\right)
\left\|Z_{k-\left\lfloor\frac{k-K}{h}\right\rfloor h}\right\|\nonumber\\
&+\sum^{\left\lfloor\frac{k}{h}\right\rfloor-1}_{l=\left\lceil\frac{K}{h}\right\rceil+1}
\prod^{\left\lceil\frac{k}{h}\right\rceil}_{q=\left\lceil\frac{K}{h}\right\rceil+\kappa+l+1}
\left(1-\frac{\lambda_{\min}(W)h^{1-p}}{q^p}\right)O\left(\frac{1}{l^{2p}}\right).\nonumber
\end{align}
Then by it and Lemma \ref{LemmaA23_WKZ2024}, we learn $\|Z_k\|=O\left(\frac{1}{k^p}\right)$, which implies
$\mathbb{E}\|\tilde{\Theta}_k\|^2 = O\left(\frac{1}{k^p}\right)$ and
$\mathbb{E}\|\varepsilon_k\|^2 = O\left(\frac{1}{k^p}\right)$.

\textbf{Part II}: The case of $b_k=\frac{1}{k}$.\\
Based on (\ref{thm_c2}), we have
\vspace{-5pt}
\begin{align}
\|Z_k\|\leq&\left(1-\lambda_{\min}(W)\sum_{l=k-h}^{k-1}\frac{1}{l+1}\right)\|Z_{k-h}\| +O\(\frac{1}{k^2}\),\nonumber\\
\leq&\left(1-\frac{\lambda_{\min}(W)h}{k}\right)\|Z_{k-h}\| +O\(\frac{1}{k^2}\),\nonumber\\
\leq&\prod^{\left\lfloor\frac{k-K}{h}\right\rfloor-1}_{l=0}\left(1-\frac{\lambda_{\min}(W)h}{k-lh}\right)
\left\|Z_{k-\left\lfloor\frac{k-K}{h}\right\rfloor h}\right\|\nonumber\\
&+\sum^{\left\lfloor\frac{k-K}{h}\right\rfloor}_{l=1}
\prod^{l-1}_{q=0}\left(1-\frac{\lambda_{\min}(W)h}{k-qh}\right)O\left(\frac{1}{(k-lh)^2}\right)\nonumber\\
\leq&\prod^{\left\lceil\frac{k}{h}\right\rceil}_{l=\left\lceil\frac{K}{h}\right\rceil+\kappa+1}
\left(1-\frac{\lambda_{\min}(W)h}{lh}\right)
\left\|Z_{k-\left\lfloor\frac{k-K}{h}\right\rfloor h}\right\|\nonumber\\
&+\sum^{\left\lfloor\frac{k}{h}\right\rfloor-1}_{l=\left\lceil\frac{K}{h}\right\rceil+1}
\prod^{\left\lceil\frac{k}{h}\right\rceil}_{q=\left\lceil\frac{K}{h}\right\rceil+\kappa+l+1}
\left(1-\frac{\lambda_{\min}(W)h}{qh}\right)O\left(\frac{1}{(lh)^2}\right)\nonumber\\
\leq&\prod^{\left\lceil\frac{k}{h}\right\rceil}_{l=\left\lceil\frac{K}{h}\right\rceil+\kappa+1}
\left(1-\frac{\lambda_{\min}(W)}{l}\right)
\left\|Z_{k-\left\lfloor\frac{k-K}{h}\right\rfloor h}\right\|\nonumber\\
&+\sum^{\left\lfloor\frac{k}{h}\right\rfloor-1}_{l=\left\lceil\frac{K}{h}\right\rceil+1}
\prod^{\left\lceil\frac{k}{h}\right\rceil}_{q=\left\lceil\frac{K}{h}\right\rceil+\kappa+l+1}
\left(1-\frac{\lambda_{\min}(W)}{q}\right)O\left(\frac{1}{l^2}\right),\nonumber
\end{align}
where $\kappa=\left\lceil\frac{k-K}{h}\right\rceil-\left\lfloor\frac{k-K}{h}\right\rfloor$.
Then by Lemma \ref{Lemma1_ZWZ2021}, we just need to prove $\lambda_{\min}(W)>1$
to ensure $\|Z_k\|=O\left(\frac{1}{k}\right)$, implying
$\mathbb{E}\|\tilde{\Theta}_k\|^2 = O\left(\frac{1}{k}\right)$ and
$\mathbb{E}\|\varepsilon_k\|^2 = O\left(\frac{1}{k}\right)$.

Similar to the proof of Theorem \ref{thm_c}, if $\gamma>\frac{\beta}{2\underline{g}\underline{\pi}\delta_{\psi}^2}
\big(2c_h\bar{N}+\sigma\(\bar{f}^2 \bar{\phi}^4+\lambda_{m}^2\) +\frac{8\beta c_h^4\bar{N}^2}{\sigma^2({3}\beta\sigma-2c_h)}+
\frac{c_h}{\beta} \big)$, we have $w_{1}>\frac{w_{2}^{2}}{w_{3}-1}+1$.
If $\beta>\frac{2c_h}{3\sigma}$, then, $w_{3}-1>0$, $w_{1}(w_{3}-1)>w_{2}^{2}+w_{3}-1$. Therefore,
\vspace{-7pt}
$$(w_{1}+w_{3})^{2}-4(w_{1}w_{3}-w_{2}^{2})<(w_{1}+w_{3}-2)^{2}.$$
\vspace{-7pt}
Hence, $\lambda_{\min}(W)>1$.  This completes this part's proof.

\vspace{-5pt}
\section{Numerical example}\label{sec:sim}

This section demonstrates the effectiveness of the proposed EFTQDI algorithm through a numerical example. Additionally, we illustrate the joint effect of the sensors: with only one-bit information exchange, the networked sensors collectively achieve an estimation task that no individual sensor could accomplish alone.

\begin{figure}[htbp]
\centering
\subfigure[Graph $\mathcal{G}^{(1)}$]
{
	\begin{minipage}{2.8cm}
	\centering
	\includegraphics[width=2.8cm]{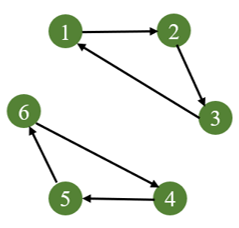}
	\end{minipage}
}	
\subfigure[Graph $\mathcal{G}^{(2)}$]
{
	\begin{minipage}{2.8cm}
	\centering
	\includegraphics[width=2.8cm]{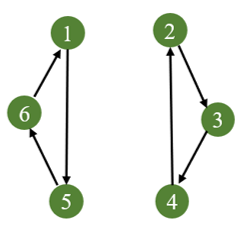}
	\end{minipage}
}
\subfigure[Graph $\mathcal{G}^{(3)}$]
{
	\begin{minipage}{2.8cm}
	\centering
	\includegraphics[width=2.8cm]{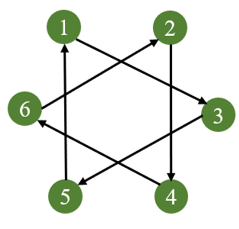}
	\end{minipage}
}	
\subfigure[Graph $\mathcal{G}^{(4)}$]
{
	\begin{minipage}{2.8cm}
	\centering
	\includegraphics[width=2.8cm]{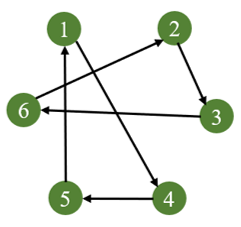}
	\end{minipage}
}
\caption{Switching communication topology} %
\label{fig1}
\end{figure}

\textbf{Example 1}
Consider a WSN with $m=6$ sensors, whose dynamics obey (\ref{M})-(\ref{est_c}) with $n=3$.
The communication graph sequence $\big\{\mathcal{G}_k\big\}$ switches among $\mathcal{G}^{(1)}$, $\mathcal{G}^{(2)}$, $\mathcal{G}^{(3)}$ and $\mathcal{G}^{(4)}$ as shown in Figure \ref{fig1}. For all $u=1,2,3,4$, $a_{i j}^{(u)}=\frac{2}{5}$ if $(j, i) \in \mathcal{E}^{(u)}$; and 0, otherwise. The graph sequence $\left\{\mathcal{G}_k\right\}$ switches according to a Markovian chain $\{m_k\}$, with initial probability $p_{u, 1}=\mathbb{P}\left\{\mathcal{G}_1=\mathcal{G}^{(u)}\right\}=$ $\frac{1}{4}$ and the following transition probability matrix:
\vspace{-5pt}
$$
\mathbf{P}=\left\{p_{u v}\right\}_{4 \times 4}=\left[\begin{array}{cccc}
{1}/{2} & {1}/{2} & 0 & 0 \\
0 & {1}/{2} & {1}/{2} & 0 \\
0 & 0 & {1}/{2} & {1}/{2} \\
{1}/{2} & 0 & 0 & {1}/{2}
\end{array}\right]
$$
where $p_{u v}=\mathbb{P}\left\{m_k=v \mid m_{k-1}=u\right\}$. Therefore, the stationary distribution $\pi_u=\frac{1}{4}$ for all $u=1,2,3,4$.

In the dynamic model (\ref{M}), the unknown parameter $\theta=[1,1,-1]^{T}$ with prior information $\theta\in\Omega=[0,2]\times[-2,0]\times[0,2]$. The observation noise $d_{k,i}$ is independent and identically distributed Gaussian with zero mean and variance $8^2$. Let $\phi_{k,i} \in \mathbb{R}^{3}$ be generated by the following state space model
\vspace{-5pt}
$$
\begin{cases}
x_{k,i}=A_i x_{k-1,i}+B_i \eta_{k,i}, \\
\phi_{k,i}=C_i x_{k,i},
\end{cases}i=1.,\ldots,6,
$$
where $x_{k,i} \in \mathbb{R}^{3}$ with $x_{0,i}=[1.3,1.3,1.3]^T $, $\eta_{k,i} \in \mathbb{R}$ is independent and identically distributed with $\eta_{k,i}$ following uniform distribution on the interval $[-0.1,0.1]$ and
\vspace{-5pt}
$$
\begin{aligned}
& A_{1}=\diag\{1,1/2,1/2\},A_{4}=\diag\{1,5/6,5/6\} \\
& A_{2}=\diag\{1/2,1,1/2\},A_5=\diag\{5/6,1,5/6\}, \\
& A_{3}=\diag\{1/2,1/2,1\},A_{6}=\diag\{5/6,5/6,1\} \\
& B_1=B_4=[1,0,0]^T,
  B_2=B_5=[0,1,0]^T,\\
& B_3=B_6=[0,0,1]^T,\\
& C_1=\diag\{1,0,0\},C_4=\diag\{-1,0,0\},\\
& C_2=\diag\{0,1,0\},C_5=\diag\{0,-1,0\},\\
& C_3=\diag\{0,0,1\},C_6=\diag\{0,0,-1\},\\
\end{aligned}
$$
It can be verified that Assumption \ref{AR} is satisfied with $h=2$. Besides, the thresholds of quantized measurements are $C_{i}=1$ for all $i=1,\ldots,6$.
In the quantized communication mechanism (\ref{est_c}), the communication noise $\omega_{k,ij}$ is independent and identically distributed Gaussian with zero mean and variance $1$, and the threshold of quantized communications is $C_{ij}=0$ for all the channel $(j,i)\in \mathcal{E}_{\mathcal{C}}$.

Then, we apply the proposed EFTQDI algorithm with two types of step sizes to produce the estimates, where the linear encoding rule $\psi_k$ in (\ref{est_z}) switches sequentially among $\{[1,0,0]^T, [0,1,0]^T,[0,0,1]^T\}$. The simulation is repeated $100$ times using the same initial values $\hat{\theta}_{0,ij}=[1/2,1/2,1/2]$ and  ${\theta}_{0,i}=[1/2,1/2,1/2]$ to calculate the empirical variance of estimation errors, representing the mean square errors.

\begin{figure}[htbp]
	\centering
	\includegraphics[width=7.6cm]{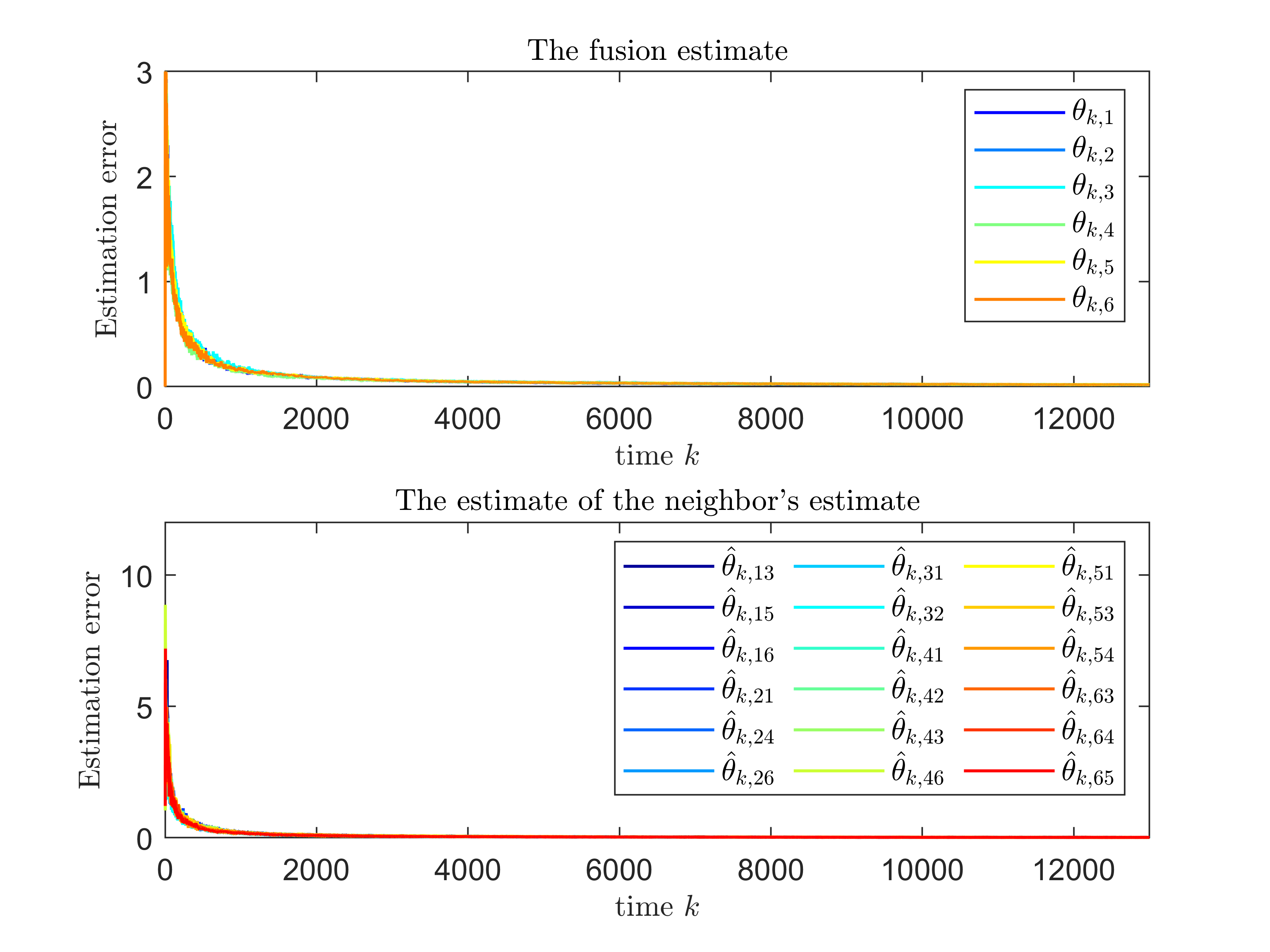}
	\caption{Convergence of the two estimates with $b_{k}=\frac{1}{k}$.}
	\label{fig_cp1}
\end{figure}

\begin{figure}[htbp]
	\centering
	\includegraphics[width=7.6cm]{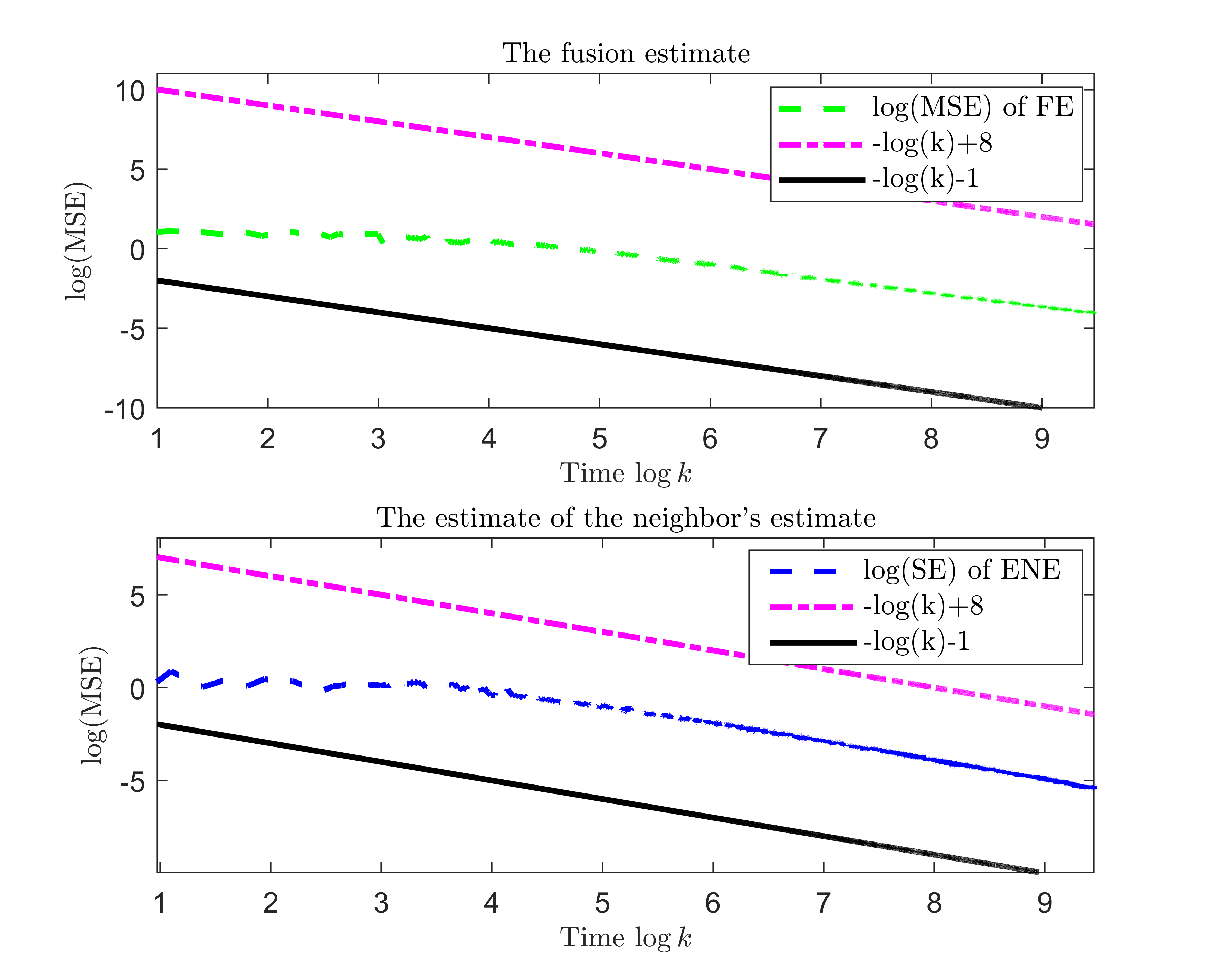}
	\caption{Convergence rate of the two estimates with $b_{k}=\frac{1}{k}$.}
	\label{fig_crp1}
\end{figure}

In the first case, the step size for the proposed EFTQDI algorithm is set as $b_{k}=\frac{1}{k}$ with the step coefficients $\beta=39$ and $\gamma=74$.
The mean square errors (MSEs) of both the fusion estimate (FE) and the estimate of the neighboring estimate (ENE) are shown in Figure \ref{fig_cp1}. 
These results indicate that the fusion estimate and ENE converge to the true parameter and fusion estimate, respectively. Additionally, Figure \ref{fig_crp1} shows a linear relationship between the logarithm of the MSEs and the logarithm of the index $k$, which indicates the mean square convergence rates of the estimation errors are $O(\frac{1}{k})$.

\begin{figure}[htbp]
	\centering
	\includegraphics[width=7.6cm]{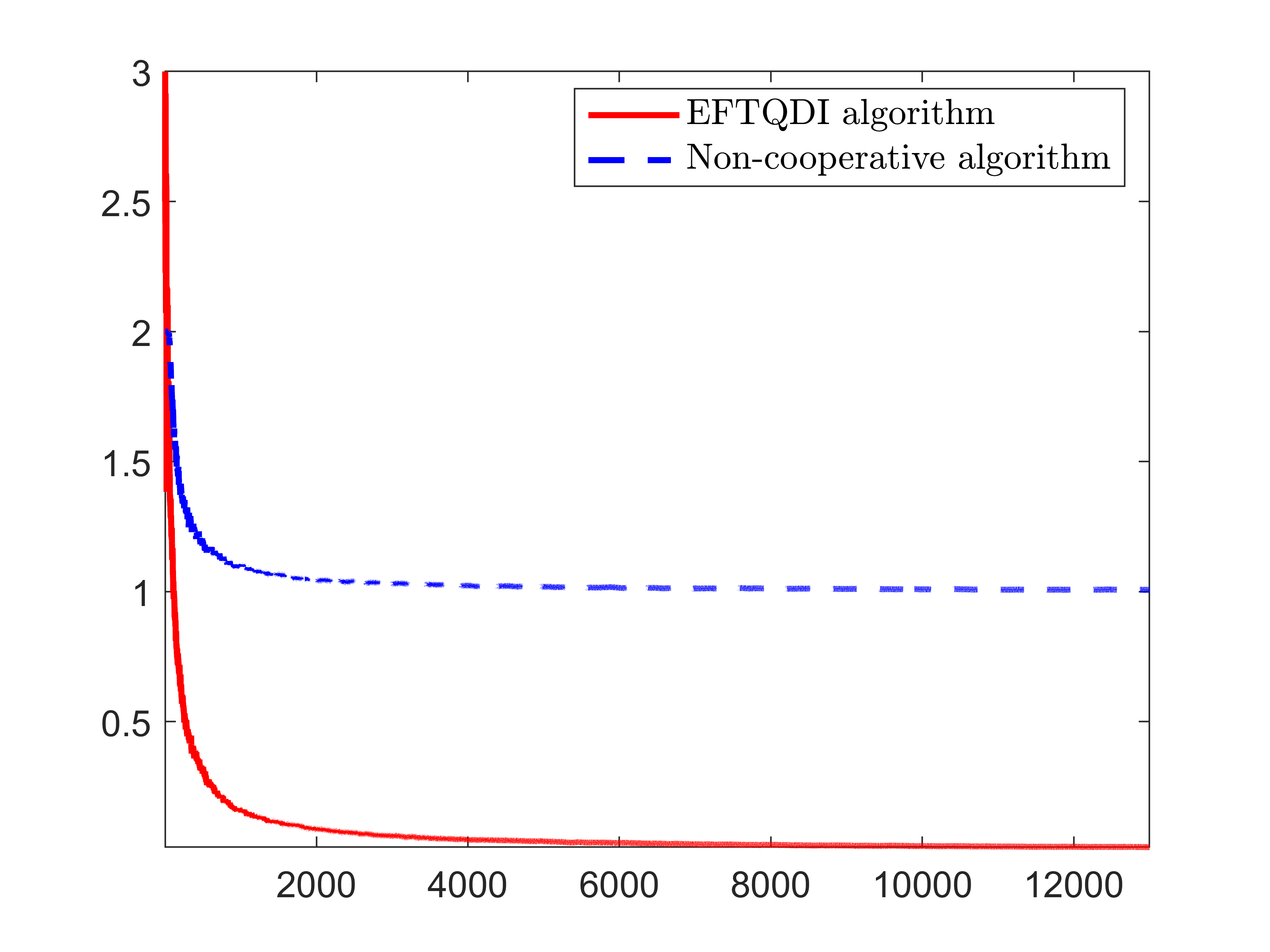}
	\caption{The comparison between the EFTQDI algorithm and non-cooperative algorithm under $b_{k}=\frac{1}{k}$.}
	\label{fig_cdbp1}
\end{figure}

Figure \ref{fig_cdbp1} shows the MSEs trajectories of the proposed EFTQDI algorithm compared to its non-cooperative counterpart (i.e., ${\theta}_{k,i}=\Pi_{\Omega}\{{\theta}_{k-1,i}+\beta b_k\phi_{k,i}(\hat{F}_{k,i}-s_{k,i})\}$).
The results indicate that, unlike the non-cooperative algorithm, which does not converge to zero, the EFTQDI algorithm's estimate converges to the true parameter. This highlights the joint effect of sensors under quantized communication, as one-bit information exchange among sensors enables the estimation task that individual sensors cannot achieve alone.

\begin{figure}[htbp]
	\centering
	\includegraphics[width=8cm]{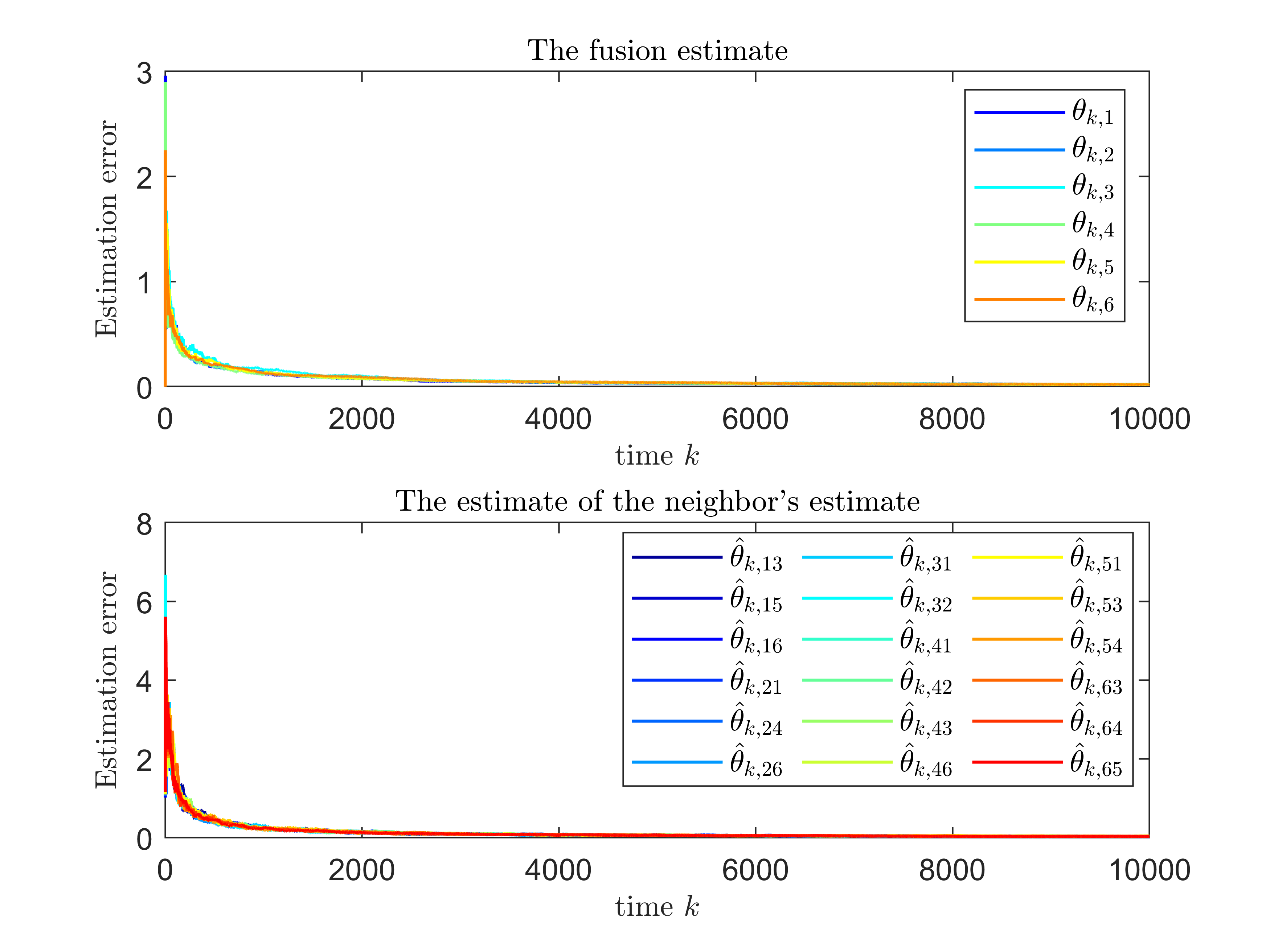}
	\caption{Convergence of the two estimates with $b_{k}=\frac{1}{k^{4/5}}$.}
	\label{fig_cp2}
\end{figure}

\begin{figure}[htbp]
	\centering
	\includegraphics[width=7.6cm]{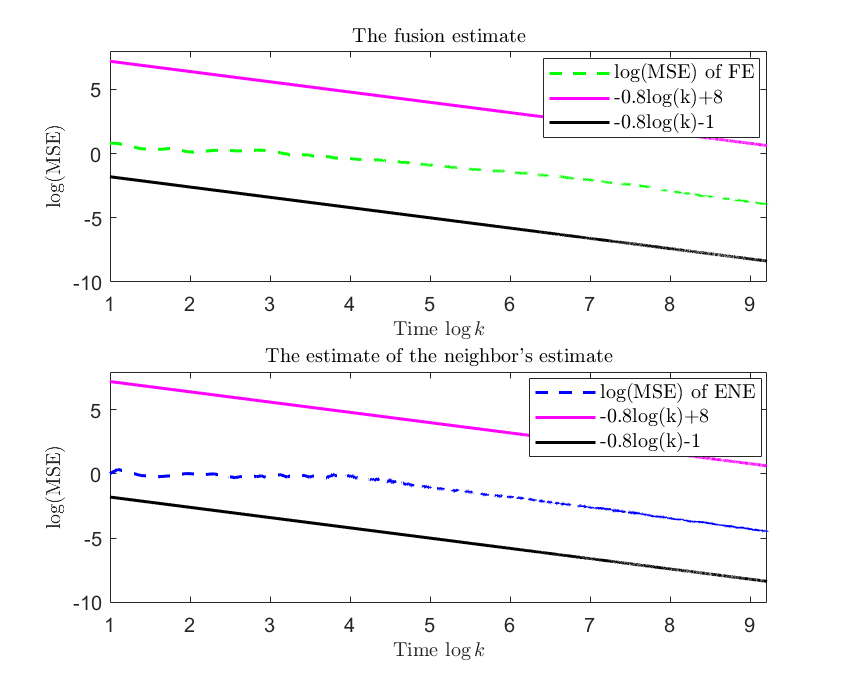}
	\caption{Convergence rate of the two estimates with $b_{k}=\frac{1}{k^{4/5}}$.}
	\label{fig_crp2}
\end{figure}

\begin{figure}[htbp]
	\centering
	\includegraphics[width=7.6cm]{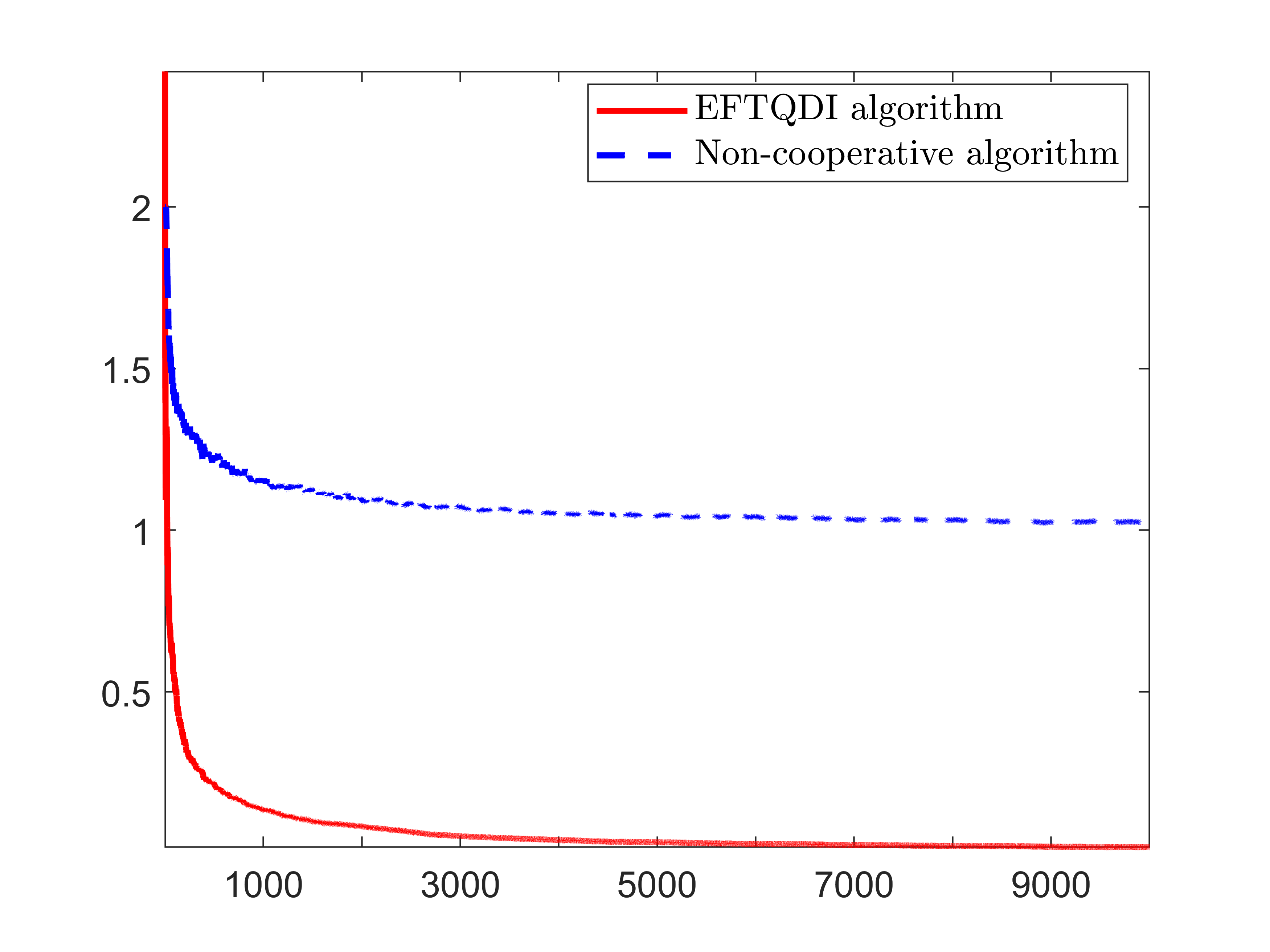}
	\caption{The comparison between the EFTQDI algorithm and non-cooperative algorithm under $b_{k}=\frac{1}{k^{4/5}}$.}
	\label{fig_cdbp2}
\end{figure}

In the second case, the step size in the EFTQDI algorithm is set as $b_{k}=\frac{1}{k^{4/5}}$ with step coefficients $\beta=16$ and $\gamma=65$.
Figure \ref{fig_cp2} shows that both the fusion estimate and the estimate of the neighboring estimate converge, respectively. Besides, Figure \ref{fig_crp2} shows that the logarithms of MSEs of the two estimates scale linearly with $\frac{4}{5}\log k$, indicating a mean square convergence rate of $O(\frac{1}{k^{4/5}})$.
Figure \ref{fig_cdbp2} shows the MSE trajectories of the EFTQDI algorithm and its corresponding non-cooperative algorithm, which demonstrates again the joint effect of the sensors under quantized communication.

\section{Conclusion}\label{sec:con}

This paper addresses distributed parameter estimation in stochastic dynamic systems with quantized measurements, focusing on quantized communication and Markov-switching directed topologies. A compression encoding method satisfying persistent excitation is introduced, ensuring original signal recovery from quantized transmitted data. Based on this, an EFTQDI algorithm is proposed, leveraging a stochastic approximation idea. This approach first estimates the neighboring estimates through quantized communication, followed by a consensus-based fusion strategy to integrate these estimates.
By constructing and analyzing two coupled Lyapunov functions for estimation errors, we establish the mean-square convergence of these two estimates under a cooperative excitation condition and the assumption that the union topology includes a spanning tree. Furthermore, we demonstrate that the mean-square convergence rate aligns with the step size order given appropriate coefficient conditions.

Future work could explore the combination of event-driven mechanisms and quantized protocol, which both can reduce communication load. Developing effective event-driven quantized transmission protocols to ensure distributed estimation with minimal data transfer remains a compelling challenge.

\bibliographystyle{elsarticle-harv}
\bibliography{mybib_QDSA}

\begin{thebibliography}{42}
\expandafter\ifx\csname natexlab\endcsname\relax\def\natexlab#1{#1}\fi
\providecommand{\url}[1]{\texttt{#1}}
\providecommand{\href}[2]{#2}
\providecommand{\path}[1]{#1}
\providecommand{\DOIprefix}{doi:}
\providecommand{\ArXivprefix}{arXiv:}
\providecommand{\URLprefix}{URL: }
\providecommand{\Pubmedprefix}{pmid:}
\providecommand{\doi}[1]{\href{http://dx.doi.org/#1}{\path{#1}}}
\providecommand{\Pubmed}[1]{\href{pmid:#1}{\path{#1}}}
\providecommand{\bibinfo}[2]{#2}
\ifx\xfnm\relax \def\xfnm[#1]{\unskip,\space#1}\fi
\bibitem[{Akyildiz et~al.(2002)Akyildiz, Su, Sankarasubramaniam and
  Cayirci}]{ASSC2002}
\bibinfo{author}{Akyildiz, I. F.}, \bibinfo{author}{Su, W.},
  \bibinfo{author}{Sankarasubramaniam, Y.}, \bibinfo{author}{Cayirci, E.},
  \bibinfo{year}{(2002)}.
\newblock \bibinfo{title}{Wireless sensor networks: a survey}.
\newblock \bibinfo{journal}{Computer Networks}, \bibinfo{volume}{38},
  \bibinfo{pages}{393--422}.
\bibitem[{Bianchi et~al.(2013)Bianchi, Fort and Hachem}]{BFH2013}
\bibinfo{author}{Bianchi, P.}, \bibinfo{author}{Fort, G.},
  \bibinfo{author}{Hachem, W.}, \bibinfo{year}{(2013)}.
\newblock \bibinfo{title}{Performance of a distributed stochastic approximation
  algorithm}.
\newblock \bibinfo{journal}{IEEE Transactions on Information Theory},
  \bibinfo{volume}{59}, \bibinfo{pages}{7405--7418}.
\bibitem[{Calamai and Mor\'{e}(1987)}]{CM1987}
\bibinfo{author}{Calamai, P. H.}, \bibinfo{author}{Mor\'{e}, J.J.},
  \bibinfo{year}{(1987)}.
\newblock \bibinfo{title}{Projected gradient methods for linearly constrained
  problems}.
\newblock \bibinfo{journal}{Mathematical Programming}, \bibinfo{volume}{39},
  \bibinfo{pages}{93--116}.
\bibitem[{Dimakis et~al.(2010)Dimakis, Kar, Moura, Rabbat and
  Scaglione}]{DKMRS2010}
\bibinfo{author}{Dimakis, A.}, \bibinfo{author}{Kar, S.},
  \bibinfo{author}{Moura, J.}, \bibinfo{author}{Rabbat, M.},
  \bibinfo{author}{Scaglione, A.}, \bibinfo{year}{(2010)}.
\newblock \bibinfo{title}{Gossip algorithms for distributed signal processing}.
\newblock \bibinfo{journal}{Proceedings of the IEEE}, \bibinfo{volume}{98},
  \bibinfo{pages}{1847--1864}.
\bibitem[{Ding et~al.(2017)Ding, Wang, Ho and Wei}]{DWDG2017}
\bibinfo{author}{Ding, D.}, \bibinfo{author}{Wang, Z.}, \bibinfo{author}{Ho,
  D.}, \bibinfo{author}{Wei, G.}, \bibinfo{year}{(2017)}.
\newblock \bibinfo{title}{Distributed recursive filtering for stochastic
  systems under uniform quantizations and deception attacks through sensor
  networks}.
\newblock \bibinfo{journal}{Automatica}, \bibinfo{volume}{78},
  \bibinfo{pages}{231--240}.
\bibitem[{Fu et~al.(2022)Fu, Chen and Zhao}]{FCZ2022}
\bibinfo{author}{Fu, K.}, \bibinfo{author}{Chen, H. F.}, \bibinfo{author}{Zhao,
  W.}, \bibinfo{year}{(2022)}.
\newblock \bibinfo{title}{Distributed system identification for linear
  stochastic systems with binary sensors}.
\newblock \bibinfo{journal}{Automatica}, \bibinfo{volume}{141},
  \bibinfo{pages}{110298}.
\bibitem[{Gan and Liu(2024)}]{GL2022}
\bibinfo{author}{Gan, D.}, \bibinfo{author}{Liu, Z.}, \bibinfo{year}{(2024)}.
\newblock \bibinfo{title}{Convergence of the distributed {SG} algorithm under
  cooperative excitation condition}.
\newblock \bibinfo{journal}{IEEE Transactions on Neural Networks and Learning
  Systems}, \bibinfo{volume}{35}, \bibinfo{pages}{7087--7101}.
\bibitem[{Ge et~al.(2020)Ge, Han, Zhang, Ding and Yang}]{GHZDY2020}
\bibinfo{author}{Ge, X.}, \bibinfo{author}{Han, Q. L.}, \bibinfo{author}{Zhang,
  X.}, \bibinfo{author}{Ding, L.}, \bibinfo{author}{Yang, F.},
  \bibinfo{year}{(2020)}.
\newblock \bibinfo{title}{Distributed event-triggered estimation over sensor
  networks: A survey}.
\newblock \bibinfo{journal}{IEEE Transactions on Cybernetics},
  \bibinfo{volume}{50}, \bibinfo{pages}{1306--1320}.
\bibitem[{Guo and Zhao(2013)}]{GZ2013}
\bibinfo{author}{Guo, J.}, \bibinfo{author}{Zhao, Y. L.}, \bibinfo{year}{(2013)}.
\newblock \bibinfo{title}{Recursive projection algorithm on {FIR} system
  identification with binary-valued observations}.
\newblock \bibinfo{journal}{Automatica}, \bibinfo{volume}{49},
  \bibinfo{pages}{3396--3401}.
\bibitem[{Han et~al.(2023a)Han, Wang, Dong, Yi and Alsaadi}]{HWDYA2023}
\bibinfo{author}{Han, F.}, \bibinfo{author}{Wang, Z.}, \bibinfo{author}{Dong,
  H.}, \bibinfo{author}{Yi, X.}, \bibinfo{author}{Alsaadi, F.},
  \bibinfo{year}{(2023)}a.
\newblock \bibinfo{title}{Distributed $h_{\infty }$-consensus estimation for
  random parameter systems over binary sensor networks: A local performance
  analysis method}.
\newblock \bibinfo{journal}{IEEE Transactions on Network Science and
  Engineering}, \bibinfo{volume}{10}, \bibinfo{pages}{2334--2346}.
\bibitem[{Han et~al.(2023b)Han, Wang, Dong, Yi and Alsaadi}]{HWDYA2023I}
\bibinfo{author}{Han, F.}, \bibinfo{author}{Wang, Z.}, \bibinfo{author}{Dong,
  H.}, \bibinfo{author}{Yi, X.}, \bibinfo{author}{Alsaadi, F.},
  \bibinfo{year}{(2023)}b.
\newblock \bibinfo{title}{Recursive distributed filtering for time-varying
  systems over sensor network via rayleigh fading channels: Tackling binary
  measurements}.
\newblock \bibinfo{journal}{IEEE Transactions on Signal and Information
  Processing over Networks}, \bibinfo{volume}{9}, \bibinfo{pages}{110--122}.
\bibitem[{Huang et~al.(2010)Huang, Dey, Nair and Manton}]{HDNM2010}
\bibinfo{author}{Huang, M.}, \bibinfo{author}{Dey, S.}, \bibinfo{author}{Nair,
  G.}, \bibinfo{author}{Manton, J.}, \bibinfo{year}{(2010)}.
\newblock \bibinfo{title}{Stochastic consensus over noisy networks with
  markovian and arbitrary switches}.
\newblock \bibinfo{journal}{Automatica}, \bibinfo{volume}{46},
  \bibinfo{pages}{1571--1583}.
\bibitem[{Kar et~al.(2012)Kar, Moura and Ramanan}]{KMR2012}
\bibinfo{author}{Kar, S.}, \bibinfo{author}{Moura, J.},
  \bibinfo{author}{Ramanan, K.}, \bibinfo{year}{(2012)}.
\newblock \bibinfo{title}{Distributed parameter estimation in sensor networks:
  Nonlinear observation models and imperfect communication}.
\newblock \bibinfo{journal}{IEEE Transactions on Information Theory},
  \bibinfo{volume}{58}, \bibinfo{pages}{3575--3605}.
\bibitem[{Lei and Chen(2020)}]{LC2020}
\bibinfo{author}{Lei, J.}, \bibinfo{author}{Chen, H. F.}, \bibinfo{year}{(2020)}.
\newblock \bibinfo{title}{Distributed stochastic approximation algorithm with
  expanding truncations}.
\newblock \bibinfo{journal}{IEEE Transactions on Automatic Control},
  \bibinfo{volume}{65}, \bibinfo{pages}{664--679}.
\bibitem[{Li et~al.(2023)Li, Wang, Lu and Xu}]{LWLX2023}
\bibinfo{author}{Li, J.}, \bibinfo{author}{Wang, Z.}, \bibinfo{author}{Lu, R.},
  \bibinfo{author}{Xu, Y.}, \bibinfo{year}{(2023)}.
\newblock \bibinfo{title}{Distributed filtering under constrained bit rate over
  wireless sensor networks: Dealing with bit rate allocation protocol}.
\newblock \bibinfo{journal}{IEEE Transactions on Automatic Control},
  \bibinfo{volume}{68}, \bibinfo{pages}{1642--1654}.
\bibitem[{Li and Zhang(2010)}]{LZ2010}
\bibinfo{author}{Li, T.}, \bibinfo{author}{Zhang, J. F.}, \bibinfo{year}{(2010)}.
\newblock \bibinfo{title}{Consensus conditions of multi-agent systems with
  time-varying topologies and stochastic communication noises}.
\newblock \bibinfo{journal}{IEEE Transactions on Automatic Control},
  \bibinfo{volume}{55}, \bibinfo{pages}{2043--2057}.
\bibitem[{Mateos and Giannakis(2012)}]{MG2012}
\bibinfo{author}{Mateos, G.}, \bibinfo{author}{Giannakis, G.},
  \bibinfo{year}{2012}.
\newblock \bibinfo{title}{Distributed recursive least-squares: Stability and
  performance analysis}.
\newblock \bibinfo{journal}{IEEE Transactions on Signal Processing},
  \bibinfo{volume}{60}, \bibinfo{pages}{3740--3754}.
\bibitem[{Polyak(1987)}]{S_Polyak1987}
\bibinfo{author}{Polyak, B.}, \bibinfo{year}{(1987)}.
\newblock \bibinfo{title}{Introduction to Optimization}.
\newblock \bibinfo{publisher}{New York: Optimization Soft-ware Inc.},
\bibitem[{Pottie and Kaiser(2000)}]{PK2020}
\bibinfo{author}{Pottie, G.}, \bibinfo{author}{Kaiser, W.},
  \bibinfo{year}{(2000)}.
\newblock \bibinfo{title}{Wireless integrated network sensors}.
\newblock \bibinfo{journal}{Communications of the ACM}, \bibinfo{volume}{43},
  \bibinfo{pages}{51–58}.
\bibitem[{Rawat et~al.(2014)Rawat, Singh, Chaouchi and Bonnin}]{RSCB2014}
\bibinfo{author}{Rawat, P.}, \bibinfo{author}{Singh, K.},
  \bibinfo{author}{Chaouchi, H.}, \bibinfo{author}{Bonnin, J.},
  \bibinfo{year}{(2014)}.
\newblock \bibinfo{title}{Wireless sensor networks: a survey on recent
  developments and potential synergies}.
\newblock \bibinfo{journal}{The Journal of Supercomputing},
  \bibinfo{volume}{68}, \bibinfo{pages}{1–48}.
\bibitem[{Ren and Beard(2005)}]{RB2005}
\bibinfo{author}{Ren, W.}, \bibinfo{author}{Beard, R.}, \bibinfo{year}{(2005)}.
\newblock \bibinfo{title}{Consensus seeking in multiagent systems under
  dynamically changing interaction topologies}.
\newblock \bibinfo{journal}{IEEE Transactions on Automatic Control},
  \bibinfo{volume}{50}, \bibinfo{pages}{655--661}.
\bibitem[{Seneta(2006)}]{S_Seneta2006}
\bibinfo{author}{Seneta, E.}, \bibinfo{year}{(2006)}.
\newblock \bibinfo{title}{Non-negative Matrices and Markov Chains}.
\newblock \bibinfo{publisher}{Springer: NY, USA}.
\bibitem[{Sinha and Griscik(1971)}]{RM1951}
\bibinfo{author}{Sinha, N.}, \bibinfo{author}{Griscik, M.},
  \bibinfo{year}{(1971)}.
\newblock \bibinfo{title}{A stochastic approximation method}.
\newblock \bibinfo{journal}{IEEE Transactions on Systems, Man, and Cybernetics},
  \bibinfo{volume}{SMC-1}, \bibinfo{pages}{338--344}.
\bibitem[{Swenson et~al.(2020)Swenson, Sridhar and Poor}]{SSP2020}
\bibinfo{author}{Swenson, B.}, \bibinfo{author}{Sridhar, A.},
  \bibinfo{author}{Poor, H.}, \bibinfo{year}{(2020)}.
\newblock \bibinfo{title}{On distributed stochastic gradient algorithms for
  global optimization}, in: \bibinfo{booktitle}{ICASSP 2020 - 2020 IEEE
  International Conference on Acoustics, Speech and Signal Processing}, pp.
  \bibinfo{pages}{8594--8598}.
\bibitem[{Taj and Cavallaro(2011)}]{TC2011}
\bibinfo{author}{Taj, M.}, \bibinfo{author}{Cavallaro, A.},
  \bibinfo{year}{(2011)}.
\newblock \bibinfo{title}{Distributed and decentralized multicamera tracking}.
\newblock \bibinfo{journal}{IEEE Signal Processing Magazine},
  \bibinfo{volume}{28}, \bibinfo{pages}{46--58}.
\bibitem[{Takahashi et~al.(2010)Takahashi, Yamada and Sayed}]{TYS2010}
\bibinfo{author}{Takahashi, N.}, \bibinfo{author}{Yamada, I.},
  \bibinfo{author}{Sayed, A.}, \bibinfo{year}{(2010)}.
\newblock \bibinfo{title}{Diffusion least-mean squares with adaptive combiners:
  Formulation and performance analysis}.
\newblock \bibinfo{journal}{IEEE Transactions on Signal Processing},
  \bibinfo{volume}{58}, \bibinfo{pages}{4795--4810}.
\bibitem[{Wang et~al.(2024)Wang, Ke and Zhang}]{WKZ2024}
\bibinfo{author}{Wang, J.}, \bibinfo{author}{Ke, J.}, \bibinfo{author}{Zhang,
  J. F.}, \bibinfo{year}{(2024)}.
\newblock \bibinfo{title}{Differentially private bipartite consensus over
  signed networks with time-varying noises}.
\newblock \bibinfo{journal}{IEEE Transactions on Automatic Control},
  \bibinfo{volume}{69}, \bibinfo{pages}{5788--5803}.
\bibitem[{Wang et~al.(2018)}]{WTZ2018}
\bibinfo{author}{Wang, T.}, \bibinfo{author}{Tan, J.}, \bibinfo{author}{Zhao, Y. L.}, \bibinfo{year}{(2018)}.
\newblock \bibinfo{title}{Asymptotically efficient non-truncated identification for {FIR} systems with binary-valued outputs}.
\newblock \bibinfo{journal}{Science China Information Sciences}, \bibinfo{volume}{61},
  \bibinfo{pages}{129208}.
\bibitem[{Wang et~al.(2003)Wang, Zhang and Yin}]{WZY2003}
\bibinfo{author}{Wang, L. Y.}, \bibinfo{author}{Zhang, J. F.},
  \bibinfo{author}{Yin, G.}, \bibinfo{year}{(2003)}.
\newblock \bibinfo{title}{System identification using binary sensors}.
\newblock \bibinfo{journal}{IEEE Transactions on Automatic Control},
  \bibinfo{volume}{48}, \bibinfo{pages}{1892--1907}.
\bibitem[{Wang and Zhang(2019)}]{WZ2019}
\bibinfo{author}{Wang, Y.}, \bibinfo{author}{Zhang, J. F.},
  \bibinfo{year}{(2019)}.
\newblock \bibinfo{title}{Distributed parameter identification of quantized
  {ARMAX} systems}, in: \bibinfo{booktitle}{Proceedings of the 38th Chinese
  Control Conference}, pp. \bibinfo{pages}{1701--1706}.
\bibitem[{Wang et~al.(2021)Wang, Zhao and Zhang}]{WZZ2021}
\bibinfo{author}{Wang, Y.}, \bibinfo{author}{Zhao, Y. L.},
  \bibinfo{author}{Zhang, J. F.}, \bibinfo{year}{(2021)}.
\newblock \bibinfo{title}{Distributed recursive projection identification with
  binary-valued observations}.
\newblock \bibinfo{journal}{Journal of Systems Science and Complexity},
  \bibinfo{volume}{34}, \bibinfo{pages}{2048--2068}.
\bibitem[{Wang et~al.(2022)Wang, Zhao, Zhang and Guo}]{WZZG2022}
\bibinfo{author}{Wang, Y.}, \bibinfo{author}{Zhao, Y. L.},
  \bibinfo{author}{Zhang, J. F.}, \bibinfo{author}{Guo, J.},
  \bibinfo{year}{(2022)}.
\newblock \bibinfo{title}{A unified identification algorithm of {FIR} systems
  based on binary observations with time-varying thresholds}.
\newblock \bibinfo{journal}{Automatica}, \bibinfo{volume}{135},
  \bibinfo{pages}{109990}.
\bibitem[{Xie and Guo(2018)}]{XG2018S}
\bibinfo{author}{Xie, S.}, \bibinfo{author}{Guo, L.}, \bibinfo{year}{(2018)}.
\newblock \bibinfo{title}{Analysis of normalized least mean squares-based
  consensus adaptive filters under a general information condition}.
\newblock \bibinfo{journal}{SIAM Journal on Control and Optimization},
  \bibinfo{volume}{56}, \bibinfo{pages}{3404--3431}.
\bibitem[{Xie et~al.(2021)Xie, Zhang and Guo}]{XZG2021}
\bibinfo{author}{Xie, S.}, \bibinfo{author}{Zhang, Y.}, \bibinfo{author}{Guo,
  L.}, \bibinfo{year}{(2021)}.
\newblock \bibinfo{title}{Convergence of a distributed least squares}.
\newblock \bibinfo{journal}{IEEE Transactions on Automatic Control},
  \bibinfo{volume}{66}, \bibinfo{pages}{4952--4959}.
\bibitem[{You(2015)}]{You2015}
\bibinfo{author}{You, K.}, \bibinfo{year}{(2015)}.
\newblock \bibinfo{title}{Recursive algorithms for parameter estimation with
  adaptive quantizer}.
\newblock \bibinfo{journal}{Automatica}, \bibinfo{volume}{52},
  \bibinfo{pages}{192--201}.
\bibitem[{Zhang et~al.(2021)Zhang, Wang and Zhao}]{ZWZ2021}
\bibinfo{author}{Zhang, H.}, \bibinfo{author}{Wang, T.}, \bibinfo{author}{Zhao,
  Y. L.}, \bibinfo{year}{(2021)}.
\newblock \bibinfo{title}{Asymptotically efficient recursive identification of
  {FIR} systems with binary-valued observations}.
\newblock \bibinfo{journal}{IEEE Transactions on Systems, Man, and Cybernetics:
  Systems}, \bibinfo{volume}{51}, \bibinfo{pages}{2687--2700}.
\bibitem[{Zhang et~al.(2022)Zhang, Zhao and Guo}]{ZZG2022}
\bibinfo{author}{Zhang, L.}, \bibinfo{author}{Zhao, Y. L.}, \bibinfo{author}{Guo,
  L.}, \bibinfo{year}{(2022)}.
\newblock \bibinfo{title}{Identification and adaptation with binary-valued
  observations under non-persistent excitation condition}.
\newblock \bibinfo{journal}{Automatica}, \bibinfo{volume}{138},
  \bibinfo{pages}{110158}.
\bibitem[{Zhang and Zhang(2012)}]{ZZ2012}
\bibinfo{author}{Zhang, Q.}, \bibinfo{author}{Zhang, J. F.},
  \bibinfo{year}{(2012)}.
\newblock \bibinfo{title}{Distributed parameter estimation over unreliable
  networks with markovian switching topologies}.
\newblock \bibinfo{journal}{IEEE Transactions on Automatic Control},
  \bibinfo{volume}{57}, \bibinfo{pages}{2545--2560}.
\bibitem[{Zhao et~al.(2023)Zhao, Zhang and Kang}]{ZZK2023}
\bibinfo{author}{Zhao, Y. L.}, \bibinfo{author}{Zhang, H.}, \bibinfo{author}{Kang,
  G.}, \bibinfo{year}{(2023)}.
\newblock \bibinfo{title}{System identification under saturated precise or
  set-valued measurements}.
\newblock \bibinfo{journal}{Science China Information Sciences},
  \bibinfo{volume}{66}, \bibinfo{pages}{112204}.
\bibitem[{Zhu et~al.(2018)Zhu, Chen, Xu, Guan, Xie and Johansson}]{ZCGXLJ2018}
\bibinfo{author}{Zhu, S.}, \bibinfo{author}{Chen, C.}, \bibinfo{author}{Xu,
  J.}, \bibinfo{author}{Guan, X.}, \bibinfo{author}{Xie, L.},
  \bibinfo{author}{Johansson, K. H.}, \bibinfo{year}{(2018)}.
\newblock \bibinfo{title}{Mitigating quantization effects on distributed sensor
  fusion: A least squares approach}.
\newblock \bibinfo{journal}{IEEE Transactions on Signal Processing},
  \bibinfo{volume}{66}, \bibinfo{pages}{3459--3474}.
\bibitem[{Zhu et~al.(2017)Zhu, Liu, Soh and Xie}]{ZLSX2017}
\bibinfo{author}{Zhu, S.}, \bibinfo{author}{Liu, S.}, \bibinfo{author}{Soh,
  Y.}, \bibinfo{author}{Xie, L.}, \bibinfo{year}{(2017)}.
\newblock \bibinfo{title}{Performance analysis of averaging based distributed
  estimation algorithm with additive quantization model}.
\newblock \bibinfo{journal}{Automatica}, \bibinfo{volume}{80},
  \bibinfo{pages}{95--101}.
\bibitem[{Zhu et~al.(2015)Zhu, Soh and Xie}]{ZSX2015}
\bibinfo{author}{Zhu, S.}, \bibinfo{author}{Soh, Y.}, \bibinfo{author}{Xie,
  L.}, \bibinfo{year}{(2015)}.
\newblock \bibinfo{title}{Distributed parameter estimation with quantized
  communication via running average}.
\newblock \bibinfo{journal}{IEEE Transactions on Signal Processing},
  \bibinfo{volume}{63}, \bibinfo{pages}{4634--4646}.

\end{thebibliography}

\end{document}